\documentclass[acmsmall]{acmart}

\usepackage{color}
\usepackage{xcolor}
\usepackage{soul}

\usepackage{float}
\usepackage{subfigure}

\usepackage{multirow}
\usepackage{longtable}
\usepackage{footnote}
\usepackage[ruled,linesnumbered]{algorithm2e}

\AtBeginDocument{%
  \providecommand\BibTeX{{%
    \normalfont B\kern-0.5em{\scshape i\kern-0.25em b}\kern-0.8em\TeX}}}

\setcopyright{acmcopyright}
\copyrightyear{2020}
\acmYear{2020}
\acmDOI{10.1145/1122445.1122456}

\acmJournal{JACM}
\acmVolume{37}
\acmNumber{4}
\acmArticle{111}
\acmMonth{8}

\begin{document}

\title{Capturing Dynamics of Information Diffusion in SNS: A Survey of Methodology and Techniques}

\author{Huacheng Li}
\email{leehc@buaa.edu.cn}
\author{Chunhe Xia}
\email{xch@buaa.edu.cn}
\affiliation{%
  \institution{School of Computer Science and Engineering, Beihang University}
  \streetaddress{No. 37 Xueyuan Road}
  \city{Haidian}
  \state{Beijing}
  \country{China}
}

\author{Tianbo Wang}
\authornote{Corresponding author is Tianbo Wang. }
\email{wangtb@buaa.edu.cn}
\authornotemark[0]
\affiliation{%
  \institution{School of Cyber Science and Technology, Beihang University}
  \streetaddress{No. 37 Xueyuan Road}
  \city{Haidian}
  \state{Beijing}
  \country{China}
}

\author{Sheng Wen}
\email{swen@@swin.edu.au}
\affiliation{%
  \institution{Department of Computer Science and Software Engineering, Swinburne University of Technology}
  \streetaddress{1 James Cook Drive}
  \city{Townsville}
  \state{Queensland}
  \country{Australia}
}

\author{Chao Chen}
\email{chao.chen@jcu.edu.au}
\affiliation{%
  \institution{College of Science and Engineering, James Cook University}
  \streetaddress{John Street}
  \city{Hawthorn}
  \state{Victoria}
  \country{Australia}
}

\author{Yang Xiang}
\email{yxiang@swin.edu.au}
\affiliation{%
  \institution{Digital Research \& Innovation Capability Platform, Swinburne University of Technology}
  \streetaddress{John Street}
  \city{Hawthorn}
  \state{Victoria}
  \country{Australia}
}

\renewcommand{\shortauthors}{Huacheng, Chunhe, and Tianbo, et al.}

\begin{abstract}
  Studying information diffusion in SNS (Social Networks Service) has remarkable significance in both academia and industry. 
  Theoretically, it boosts the development of other subjects such as statistics, sociology, and data mining.
  Practically, diffusion modeling provides fundamental support for many downstream applications (\textit{e.g.}, public opinion monitoring, rumor source identification, and viral marketing.) 
  Tremendous efforts have been devoted to this area to understand and quantify information diffusion dynamics. 
  This survey investigates and summarizes the emerging distinguished works in diffusion modeling. 
  We first put forward a unified information diffusion concept in terms of three components: information, user decision, and social vectors, followed by a detailed introduction of the methodologies for diffusion modeling. 
  And then, a new taxonomy adopting hybrid philosophy (\textit{i.e.,} granularity and techniques) is proposed, and we made a series of comparative studies on elementary diffusion models under our taxonomy from the aspects of assumptions, methods, and pros and cons. 
  We further summarized representative diffusion modeling in special scenarios and significant downstream tasks based on these elementary models. 
  Finally, open issues in this field following the methodology of diffusion modeling are discussed. 
\end{abstract}

  


\maketitle

%---------------------------------------------------------------
% SECTION 1 INTRODUCTION
%---------------------------------------------------------------

\section{Introduction}
\label{sec:Introduction}
The vigorous development of SNS (Social Network Service) has shifted the way humans use the Internet from simple information retrieval to the construction and maintenance of online social relationships, as well as the creation, communication, and sharing of online information. 
SNS seeps into every aspect of social life and has a profound impact on it. 
According to the statistical report published by official institutes\cite{cnnic.net.cn}, the number of SNS users in China reached 762 million in 2020, accounting for 95.6\% of Internet users. 
The number of Facebook's monthly active users hits 2 billion, doubled in less than five years \cite{facebook_techcrunch}. 
Millions of users make SNS powerful in many fields (\textit{e.g.}, business marketing, social governance). 
However, the risks of social networks cannot be neglected. 
Scholars have proven that fake news spread faster and wider than real news in social networks \cite{vosoughi2018the}. 
If malicious people utilize social networks to spread harmful information such as terrorism and rumors, it will cause a massive menace to social stability, such as Chinese salt scramble in 2011 \cite{chinese_salt_scramble}, Arab Spring revolution \cite{khondker2011role}, and the manipulation of political elections \cite{Badawy2018Analyzing}. 

Facing the opportunities and challenges brought by SNS, many scientists are devoted to studying the information diffusion in SNS, and information diffusion modeling is the basis of these studies. 
Information diffusion models aim at capturing the dynamics of information diffusing in social networks. 
It is of great significance to model the information diffusion process in both academia and practice. 
Academically, diffusion modeling involves multiple disciplines (e.g., statistics, complex networks, sociology, machine learning). 
Moreover, diffusion-related data itself is easily accessible, which in turn promotes the development of related disciplines. 
Practically, they are the basis of many downstream applications such as popularity prediction, influence maximization (IM), source identification, network inference, and social recommendation (see details in Section \ref{sec:DAOI:Applications}). 
Employing these applications in various social tasks (\textit{e.g.}, marketing, rumor source identification, and trending topics detection) makes society run more efficiently. 

Overall, there are two main schools in the information diffusion modeling history: \textbf{time-series} and \textbf{data-driven}. 
For the \textbf{former}, researchers analyze the data to summarize the diffusion laws, and then use explicit mathematical expressions to model the dynamics of information diffusion over time. 
They mainly include difference/differential-based and stochastic-based models for \textit{volume} prediction, progressive and non-progressive for \textit{individual adoption} prediction, and likelihood maximization models for \textit{propagation relationship} prediction. 
Each part of these models, including their input and output, has a clear physical meaning, which is the primary approach of classical diffusion modeling. 
For the \textbf{latter}, researchers expect that machine learning (ML) algorithms can learn diffusion laws from data to capture diffusion dynamics. 
Benefiting from the explosion of data resources and computing power, these models have been developed considerably. 
Especially, with the development of related technologies such as Natural Language Processing (NLP) and Reasoning, content semantics that difficult for traditional model processing can be directly learned from data, which means that a new era for information diffusion modeling is heralding. 

Although there are several surveys \cite{wang2013modelingsurvey, jiang2017identifying, li2018influencesurvey, gao2019taxonomy, guille2013informationsurvey, yao2015diffusionsurvey, singh2018survey, li2017survey, 10.1145/3161603, gao2019taxonomy, 10.1145/3433000} on information diffusion, this survey is distinct in the following aspects. 
\textbf{Fisrt}, surveys \cite{wang2013modelingsurvey, jiang2017identifying, li2018influencesurvey} focus on the downstream applications of diffusion models (\textit{e.g.}, worm propagation, source identification, IM, and popularity prediction) rather than introducing underlying diffusion models. 
Surveys \cite{guille2013informationsurvey, yao2015diffusionsurvey, singh2018survey, gao2019taxonomy, 10.1145/3433000, li2017survey} 
do not involve relationship inference models (see details in Section \ref{sec:TDM:RelationshipInference}) and non-progressive models (see details in Section \ref{sec:TDM:IAM:NonProgressive}) models. 
\textbf{Second}, the commonly used classification criteria (\textit{e.g.}, predictive/explanatory \cite{guille2013informationsurvey}, topology-based/non-topology-based \cite{guille2013informationsurvey,li2018influencesurvey}) are no longer applicable, because emerging models blur the original classification boundary.
For instance, an explanatory model \cite{gao2017novel}, which embeds information and users into a unified Euclidean semantic space, can predict the future diffusion trajectory based on geometrical relationships among information contents and users. 
Some classical non-topology-based models \cite{PDE_wang2012diffusive, he2016cost} that introduce topology factors. 
Obviously, previous classification criteria may narrow readers' thoughts. 
\textbf{Third}, the information diffusion process is driven by users' social behaviors. 
Information, social vectors, and users are an interactional and indivisible whole. 
However, the majority of existing models only focus on information diffusion actions and ignore other parts. 
With the development of NLP and other related technologies, information diffusion modeling has great academic and practical potentialities. 
\textbf{Finally}, although data-driven models can automatically learn the diffusion characteristics, they hardly introduce the knowledge of classical (Time-series) modeling techniques. Introducing the knowledge of time-series techniques can help model optimization, such as reducing the search space. Some scholars are already doing this meaningful work \cite{cao2017deephawkes, DBLP:conf/sigir/ChenZ0TZZ19, bourigault2016representation, wu2019neural}. 
Therefore, we have summarized the typical extension roadmaps of time-series models, hoping to raise readers’ inspiration. 

This survey consists of four main parts. 
First, we present preliminary and research methodology, including elementary notations, the unified diffusion modeling concept, and basic processes of information diffusion modeling. 
Second, we propose a new taxonomy of elementary diffusion models and classify them into three categories: \emph{Diffusion Volume Models}, \emph{Individual Adoption Models}, and \emph{Relationship Inference Models}. 
Besides, comparative studies on assumptions, methods, as well as pros and cons are organized by these categories to help readers acquire comprehensive knowledge of information diffusion models. 
Third, we summarized representative diffusion modeling in special scenarios as well as downstream applications based on these elementary models.
Last, we discuss open issues following the methodology of diffusion modeling. 

We summarize our analysis of existing models below: 
\begin{itemize}
  \item Information diffusion scenario can be decomposed into three components: information, user, and topology. 
  The information diffusion process is not isolated, as the three layers interact with each other. 
  However, current research does not unify these three components.  
  \item Existing diffusion models mainly describe the diffusion process from three levels of granularity: volume, individual, and propagation relationship. 
  Under the granularity-oriented taxonomy, each type of model is classified into two categories: time-series and data-driven. 
  \item Time-series individual adoption models (see details in Section \ref{sec:TDM:IAM:Progressive} and Section \ref{sec:TDM:IAM:NonProgressive}) can not be used for propagation relationship inference directly, because they are the models with many-to-many propagation relationships. They can infer the propagation probabilities between users by likelihood maximization, and select corresponding relationships to construct a propagation network. 
  \item Data-driven models avoid the intrinsic process of devising precise mathematical diffusion models, making them easier to utilize new features. 
  However, their difficulties are extracting representative and comprehensive features. 
  \item These two approaches should be mutually reinforcing. Time-series models can provide inspiration for the design of deep learning models, and the Data-driven model can explore new diffusion characteristics to improve the mathematical expression of Time-series models. 
\end{itemize}

The rest of the article is organized as follows. 
Section \ref{sec:ResearchMethodologyPreliminary} introduces the methodology and basic knowledge used in this article. 
We further provide a thorough review of elementary information diffusion models in Section \ref{sec:TaxonomyDiffusionModels}. 
Section \ref{sec:ScenarioAndApplication} summarizes and discusses some scenario-specified diffusion modeling and downstream applications. 
Section \ref{sec:Challenges} discuss open issues of diffusion modeling.
Finally, we present the conclusion in Section \ref{sec:Conclusion}.

%---------------------------------------------------------------
% SECTION 2 RESEARCH AND METHODOLOGY AND PRELIMINARY
%---------------------------------------------------------------

\begin{figure}[h]
  \centering
  \includegraphics[width=5in]{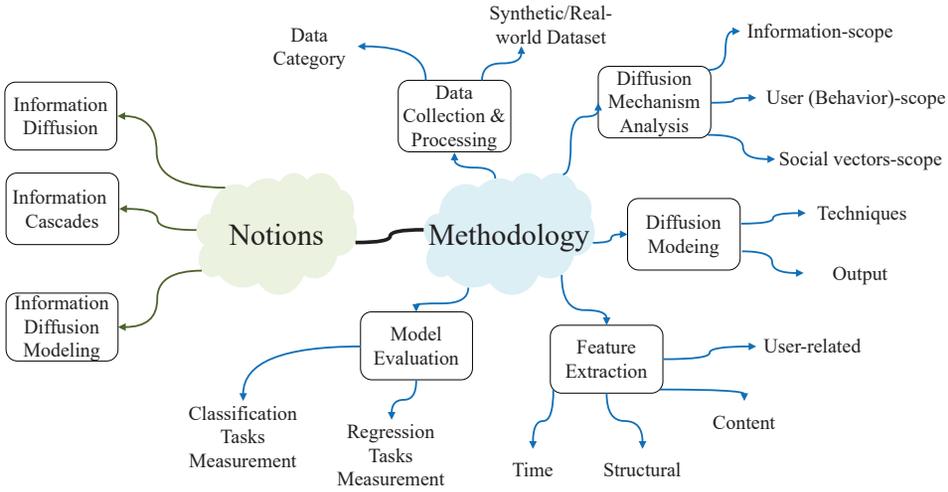}
  \caption{The skeleton of Section \ref{sec:ResearchMethodologyPreliminary}.}
  \label{FIG:Section2}
\end{figure}

\section{Preliminary and Methodology of Information Diffusion}
\label{sec:ResearchMethodologyPreliminary}
We introduce the notions and methodology of information diffusion modeling to help readers gain a rudimentary knowledge of it. 
For readability, the skeleton of this section is shown in Fig. \ref{FIG:Section2}. 

\begin{figure}
  \centering
  \subfigure[Illustration of information diffusion in social networks]{\label{FIG:DiffusionScene} \includegraphics[width=5.5in]{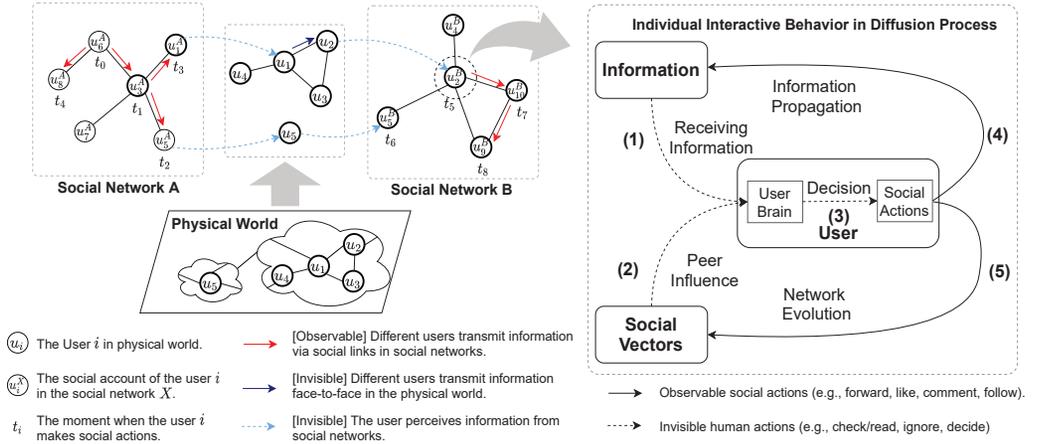}}
  \subfigure[Research methodology of diffusion modeling]{\label{FIG:Methodology} \includegraphics[width=5.5in]{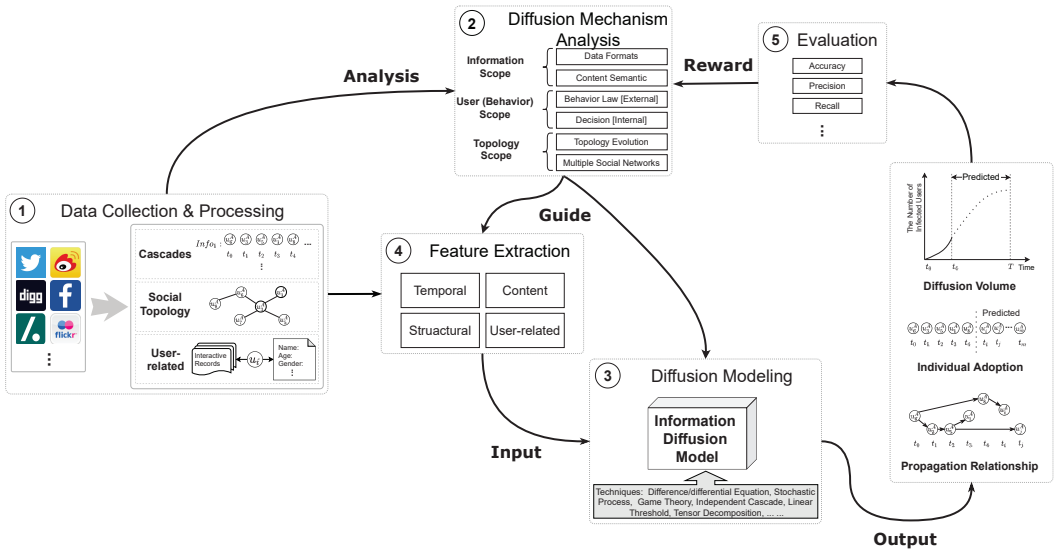}}
  \caption{Illustrations of information diffusion process and diffusion modeling methodology. (a) describes the information diffusion in social networks. The left part illustrates the information can not only be propagated via social links in single SNS-based networks, but also "jump" across multiple SNS-based networks with the help of the physical world friendships. The right part illustrates users' interactive behaviors in the information diffusion process. We can capture the information diffusion dynamics by monitoring observable social actions.  (b) describes the research methodology of information diffusion modeling. The methodology is iterative and can be summarized as five steps: data collection and processing, diffusion mechanism analysis, diffusion modeling, feature extraction, and evaluation. After collecting data, we analyze the diffusion mechanisms from three scopes: information, user (behavior), and social vectors. They guide the following diffusion modeling and feature extraction. The model performance is evaluated and then fed back to the mechanism analysis. }
  \label{FIG:DiffusionSceneAll}
\end{figure}

\subsection{Notions}
\label{sec:RMP:Notions}
\subsubsection{Information Diffusion}
Fig. \ref{FIG:DiffusionScene} is an abstract representation of the information diffusion process in social networks, and it has the following facts: 
\begin{itemize}
  \item \textit{Three components in diffusion process: information, users (or user behavior), and social vectors (i.e., transmission media)}. 
  Social vectors consist of users with their social friendships. 
  \item \textit{Diffusion process is unified.} Information, users, and social vectors influence each other in the diffusion process. 
  As shown in the right part of Fig. \ref{FIG:DiffusionScene}, users perceive information and others' opinions from their social neighbors or other channels, and then decide how to interact with it. After that, their social actions drive information propagation (e.g., retweet, comment) and social network evolution (e.g., follow, unfollow).  
  \item \textit{Information content affects diffusion process.} For a particular piece of information, acceptance of different users' is various. For instance, boys who play basketball are obviously more likely to share NBA-related messages than girls who do not.  
  \item \textit{Individual user social behaviors are diversified.} The dynamics of information diffusion can only be captured by users' social actions (action 4 and 5 in the right part of Fig. \ref{FIG:DiffusionScene}), and the interval between social actions are not fixed (timestamps in the left part of Fig. \ref{FIG:DiffusionScene}). In addition, the information can not only be transmitted via social links in single social networks, but also "jump" across multiple social networks with the help of the physical world friendships (diffusion path $u_1^A$ $\rightarrow$ $u_1$ $\rightarrow$ $u_2$ $\rightarrow$ $u_2^B$ in the left part of Fig. \ref{FIG:DiffusionScene}). Moreover, one user may share the same information to more than one social networks (diffusion path $u_5^B$ $\rightarrow$ $u_5$ $\rightarrow$ $u_5^B$ in the left part of Fig. \ref{FIG:DiffusionScene}). 
  \item \textit{Social vectors are heterogeneous.} One user can possess multiple social accounts, and friends in the physical world may not have social links in SNS-based networks. Furthermore, different SNS-based networks have various communication ways. For example, Twitter users can browse all the public information, while WhatsApp users can only access their friends. 
\end{itemize}

Here we have to distinguish meanings of these words: `\textbf{propagation},' `\textbf{transmission},' `\textbf{diffusion},' and `\textbf{dissemination}.' 
\textbf{Propagation} and \textbf{transmission} are usually used to describe information transmitted between two users, while \textbf{diffusion} and \textbf{dissemination} refer to the information spreading to other peoples. 
Therefore, we use propagation probability and transmission rate instead of diffusion probability or diffusion rate to denote the possibility that information is transmitted between two users.

\subsubsection{Information Cascades}
According to \cite{guille2013informationsurvey}, all users who share the message $Info_i$ constitute the cascade $c_i$, and their motivation can be either their interests or peer influence. 
The cascades $\mathcal{C}$ consists of many single cascade $c_i$. 
It records the information contents as well as the user's interactive behavior and time. 
As shown in the ``Data Collection $\&$ Processing'' of Fig. \ref{FIG:Methodology}, the user \(u_6^A\) is the original poster of $Info_1$ at time $t_0$. 
Then users $u_3^A$, $u_5^A$, $u_1^A$, and $u_8^A$ retweet the $Info_1$ at time $t_1$, $t_2$, $t_3$, and $t_4$, respectively. 
Notable, the interval between two adjacent time ticks may be different. 

\subsubsection{Diffusion Modeling}
For computation convenience, researchers usually regard users as pure nodes, and use ``inactive/susceptible'' and ``active/infected'' to distinguish whether a user receives the information. 
Symbols commonly used in this paper are listed in Table \ref{TAB:BasicSymbol}.

Given a social network as \(G(V, E, W)\), with user-set \(V\), social link set \(E\), and edge weight set $W$, information diffusion model \(\mathcal{M}\) aims at capturing information diffusion dynamics in social networks. 
For a specific message \(I_{message}\), only some of users \(\mathcal{I}(0) \subseteq V\) adopt the message \(I_{message}\) initially, denoted as seed users \(\mathcal{I}(0)\). 
As time elapses, the message \(I_{message}\) spread out by seed users through social links randomly, and the model \(\mathcal{M}\) captures this process.

\begin{table}
  \caption{Definition of Basic Parameters}
  \label{TAB:BasicSymbol}
  \centering
  \scriptsize
  \begin{tabular}{p{1.5cm}|p{10cm}}
      \toprule
      \textbf{Symbols}           & \textbf{Definations} \\
      \midrule
      \(G(V,E,W)\)              & social network topology with the user set $V$, the edge set $E$, and the edge weight set $W$ \\
      \(\mathcal{C}\)           & cascade set\\
      \(c\)                     & a piece of cascade\\
      \(w_c(t)\)                & weight of cascade \(c\) \\ 
      \(T\)                     & last time \(t\) of observable cascade \\
      \(t_k^c\)                 & infection time of $k$-th forwarding in the cascade $c$ \\
      \(u_k^c\)                 & the $k$-th forwarding user in the cascade $c$ \\
      %\(S\)                    & susceptible (or inactive, or un-informed) user/user\\
      %\(I\)                    & infected (or active, or informed) user/user\\
      \(N(t)\)                  & count of users at time \(t\)\\
      \(S(t)\)                  & count of susceptible (or inactive) users at time \(t\)\\
      \(I(t)\)                  & count of infected (or active) users at time \(t\)\\
      \(\mathcal{N}(t)\)        & total population of available users in social network at time \(t\)\\
      \(\mathcal{S}(t)\)        & susceptible (or inactive) user set at time \(t\)\\
      \(\mathcal{I}(t)\)        & infected (or active) user set at time \(t\) \\
      \(\Delta I(t)\)           & count of newly infected (or activated) user at time \(t\) \\
      \(\Delta\mathcal{I}(t)\)  & newly infected (or activated) user set at time \(t\)\\
      \(D_{u}^{in}(t)\)         & incoming neighbor set of user \(u\) at time $t$\\
      \(D_{u}^{out}(t)\)        & outcoming neighbor set of user \(u\) at time $t$\\
      \(D_{u}(t)\)              & neighbor set of user \(u\) at time \(t\)\\
      \(p_{uv}^{c}(t)\)         & propagation probability or transmission rate of cascade \(c\) from user \(u\) to \(v\) at time \(t\)\\
      \(p_{uv}(t)\)             & propagation probability or transmission from user \(u\) to \(v\) at time \(t\)\\ 
      \textbf{A}$(t)$           & matrix of transmission rate at time $t$ \\
      $d_u^{in}(t)$             & in-degree of the user $u$ at time $t$ \\
      $d_u^{out}(t)$            & out-degree of the user $u$ at time $t$ \\
      \(\textbf{f}_{u}^{(m)}\)  & feature vector of user \(u\) about the topic/message\(m\) \\
      \bottomrule
  \end{tabular}
\end{table}

\subsection{Research Methodology}
\label{sec:RMP:Methodology}

As shown in Figure. \ref{FIG:Methodology}, the process of information diffusion modeling is iterative, and it can be divided into five steps: Data Collection \(\&\) Processing, Diffusion Mechanism Analysis, Diffusion Modeling, Feature Extraction, and Model Evaluation. 
These five steps are described as follows:

\begin{figure}
  \subfigure[Facebook-RW]{\label{FIG:Facebook_Real} \includegraphics[width=1.2in]{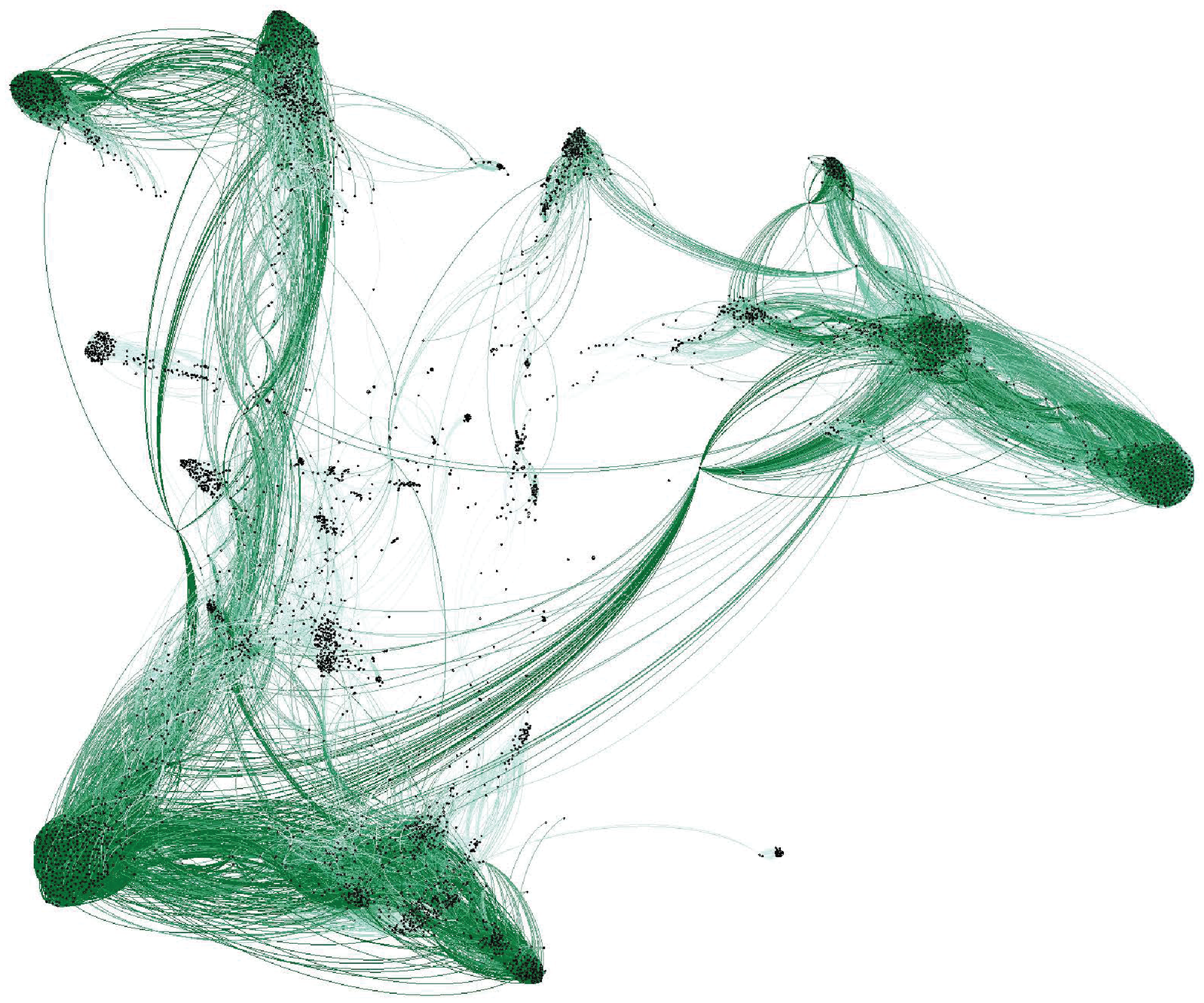}}
  \subfigure[Facebook-BTER]{\label{FIG:Facebook_BTER} \includegraphics[width=1.2in]{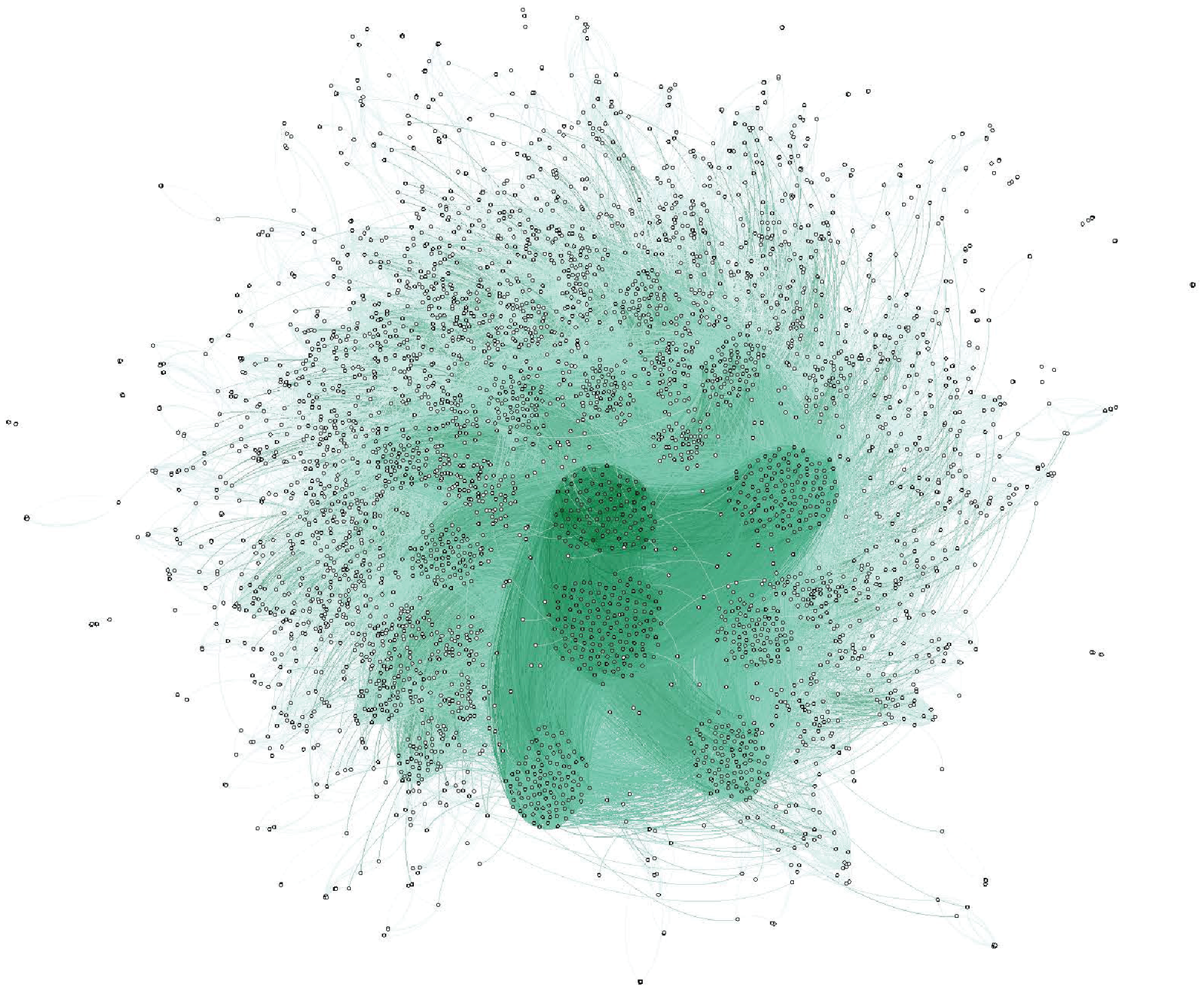}}
  \subfigure[Facebook-ER]{\label{FIG:Facebook_ER} \includegraphics[width=1.2in]{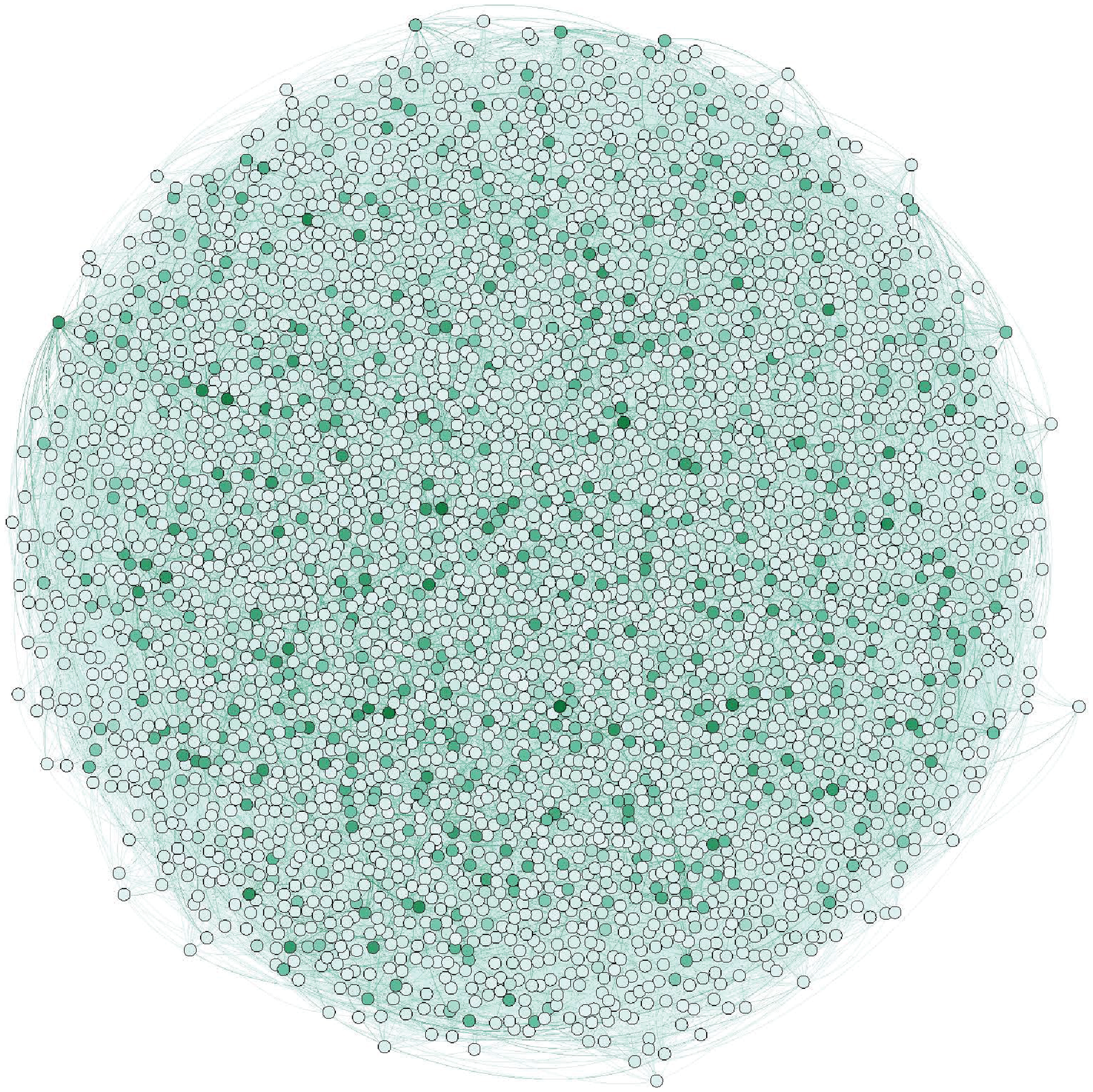}}
  \subfigure[Degree distribution of Facebook]{\label{FIG:Facebook_DegDis} \includegraphics[width=1.6in]{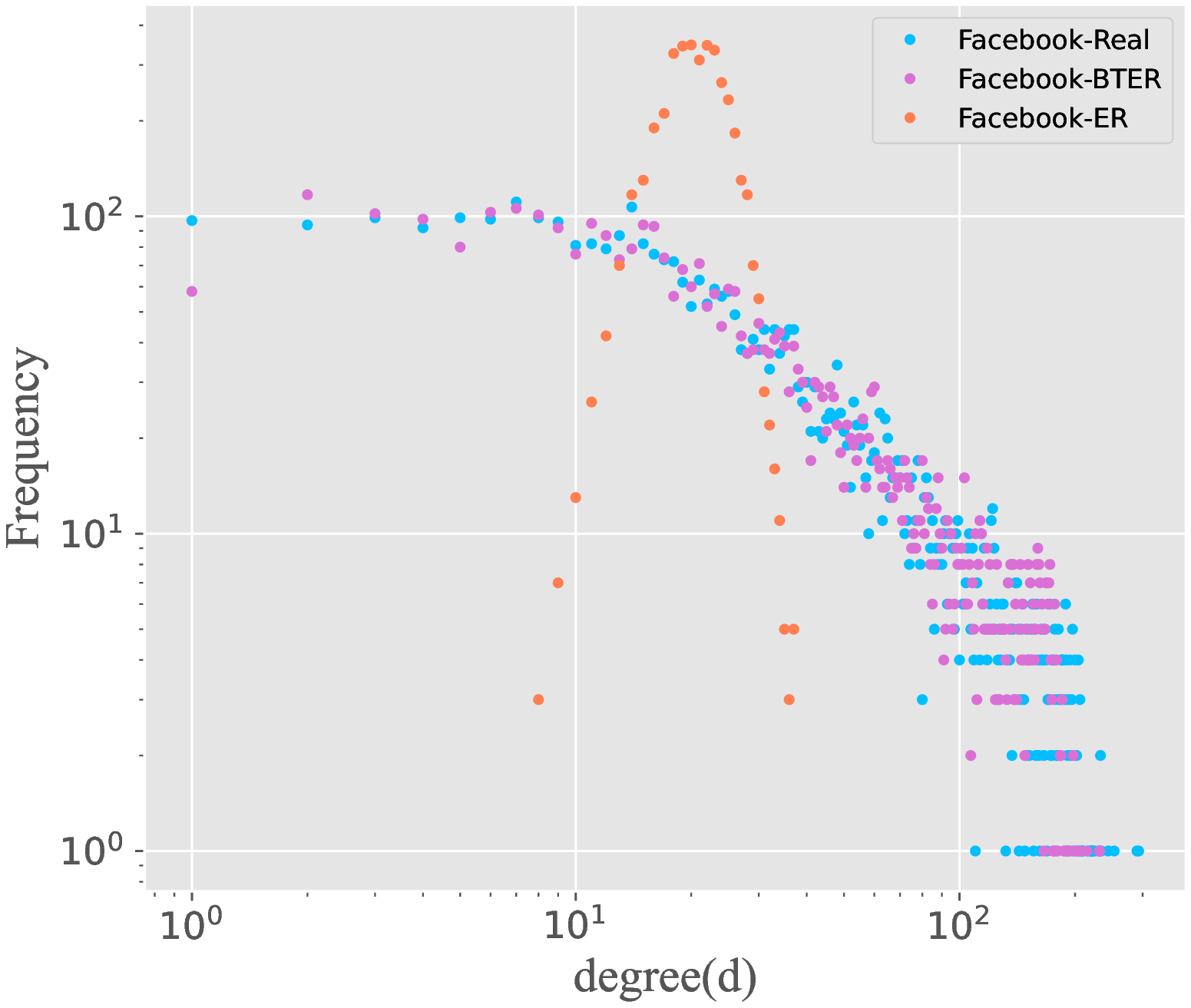}}
  \subfigure[Arxiv-RW]{\label{FIG:Arxiv_Real} \includegraphics[width=1.2in]{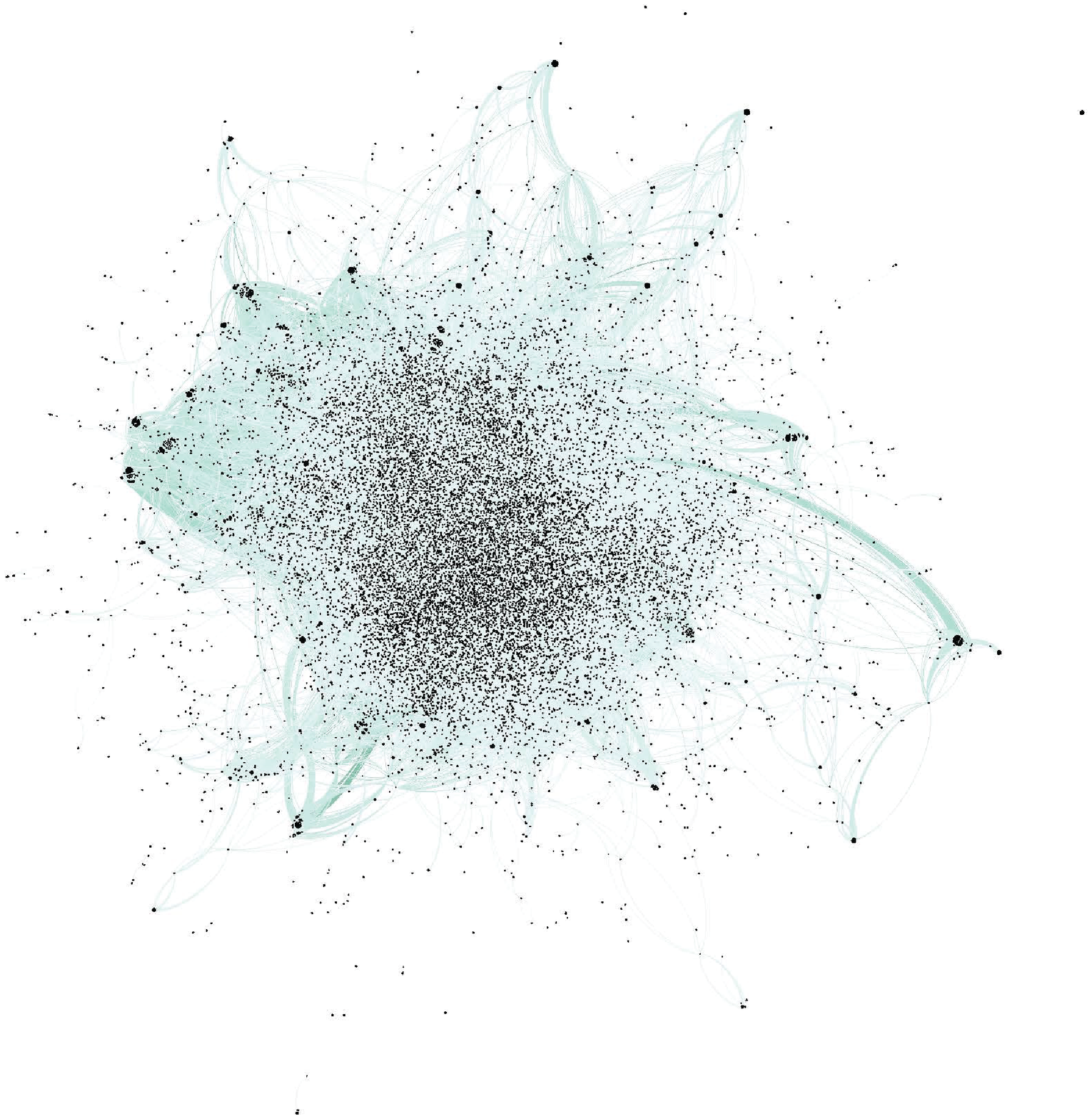}}
  \subfigure[Arxiv-BTER]{\label{FIG:Arxiv_BTER} \includegraphics[width=1.2in]{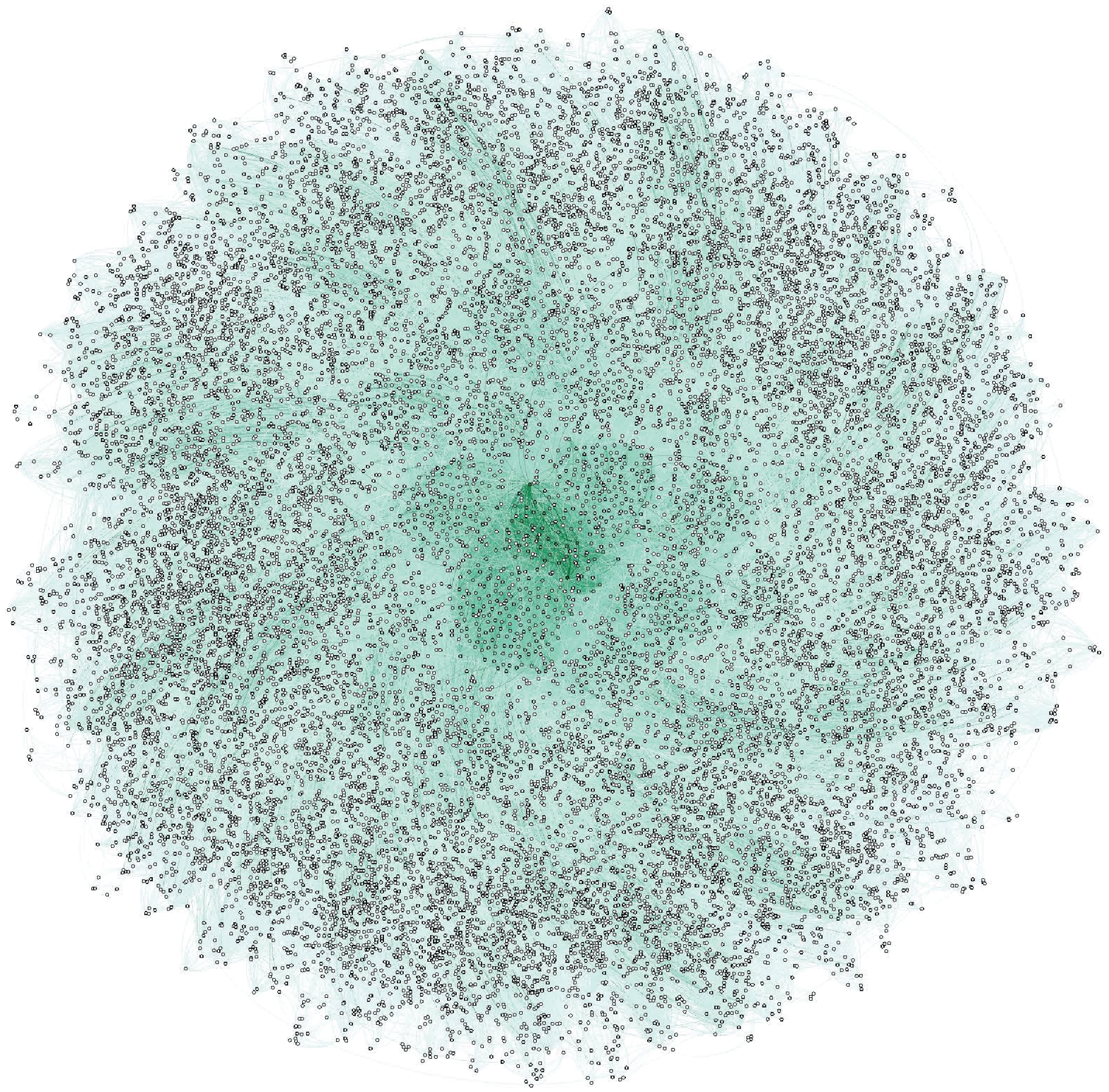}}
  \subfigure[Arxiv-ER]{\label{FIG:Arxiv_ER} \includegraphics[width=1.2in]{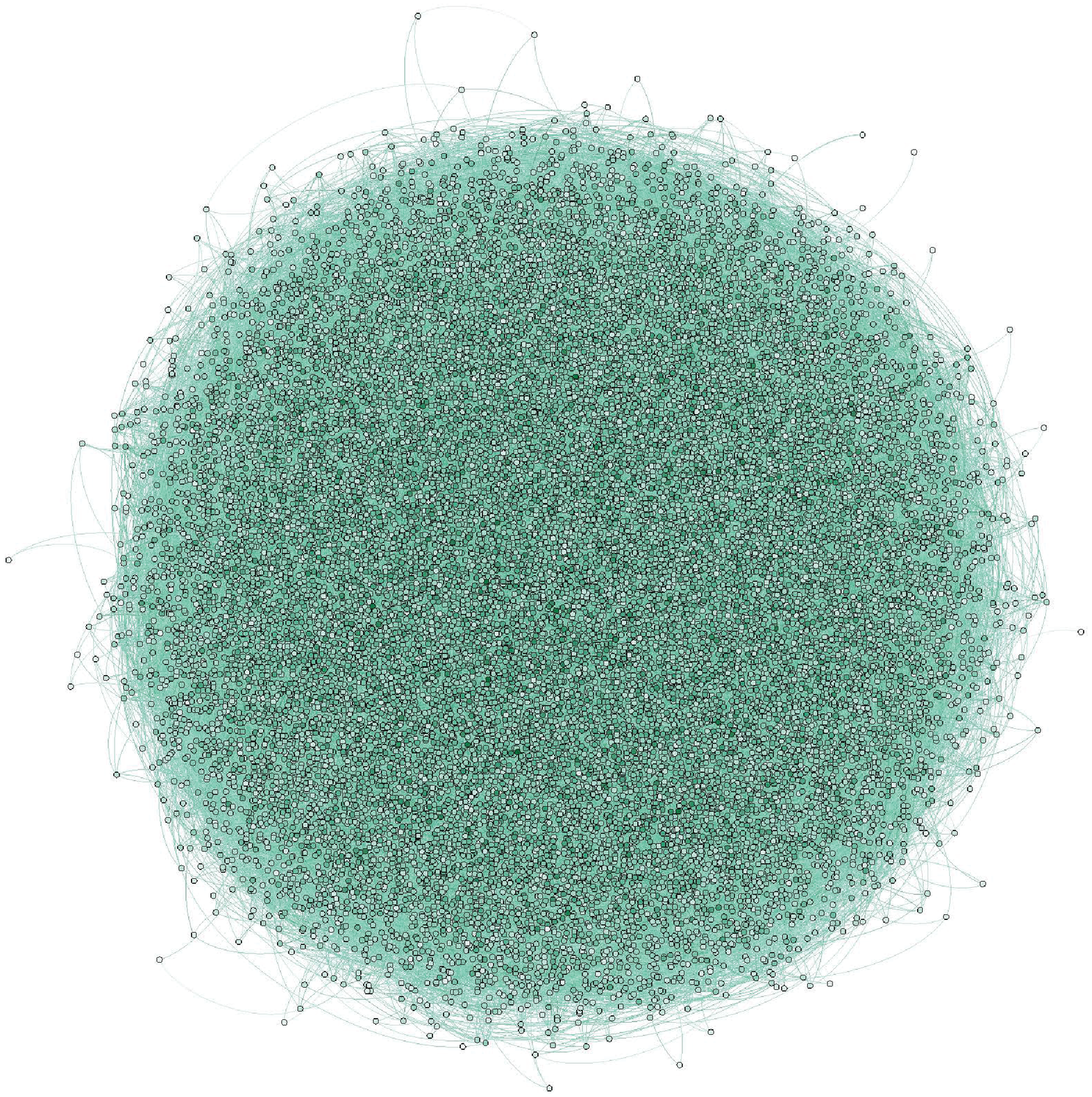}}
  \subfigure[Degree distribution of Arxiv]{\label{FIG:Arxiv_DegDis} \includegraphics[width=1.6in]{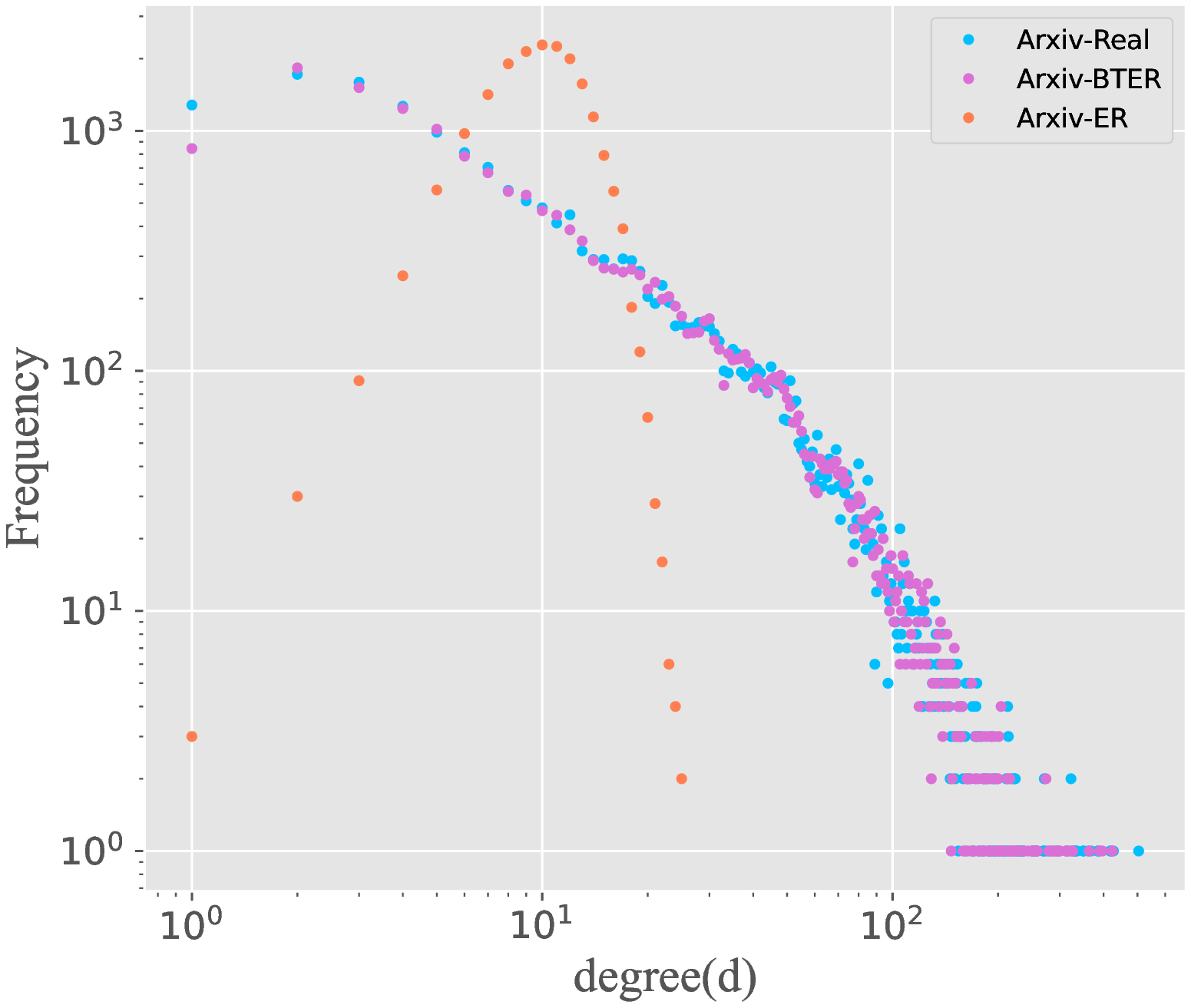}}
  \caption{Comparison of real-world and synthetic data. (a) Real-world Facebook dataset downloaded from SNAP \cite{10.5555/2999134.2999195}. (b) Block Two-Level Erdos-Renyi graph of Facebook generated by DARPA GRAPHS \cite{PhysRevE.85.056109, doi:10.1137/130914218, doi:10.1137/1.9781611972832.2, kolda2014feastpack}. (c) Erdos-Renyi graph of Facebook generated by networkx (Python package for network analysis). (d) Degree distributions of real-world, Block Two-Level Erdos-Renyi, Erdos-Renyi graphs. (e) Real-world Arxiv collaboration network dataset downloaded from SNAP \cite{10.1145/1217299.1217301}. The meanings of (f), (g), (h) are similar to (b), (c), (d). }
  \label{FIG:RealAndSynthetic}
\end{figure}

\subsubsection{Data Collection \(\&\) Processing} 
The first step of research is to obtain sufficient data. 
Social data can be processed into three categories: \emph{cascades}, \emph{social topology}, and \emph{user-related data}. 
\emph{Cascades} records the diffusion dynamics of mass social information. 
It is the ground truth of the diffusion process, and we can evaluate the model performance based on it. 
\emph{Social topology} consists of users with their social relationships. 
Evidence shows that it affects the range and speed of information diffusion \cite{DBLP:conf/icwsm/YangC10, PhysRevE.102.052316}. 
Commonly, they are major elements in predicting the information diffusion process.  
\emph{User-related data} includes user-profiles and their interactive social records. 
These data can be used to infer user preference. 
Combined with the information content, the user's interest in the information can be calculated, and the range of candidate users can be further narrowed, thereby improving the prediction accuracy. 
With the development of Deep Learning and NLP technologies, as well as the growth of computing power and data resources, this type of data has great potential in diffusion modeling. 
Notably, these three types of data must be \textbf{time-related}, because the time factors in the data can reflect the dynamics of information dissemination. 

Datasets can be categorized into either \emph{synthetic} and \emph{real-world} data. 
The \emph{synthetic} data are generated by programs following social analysis principles (\textit{e.g.}, power-law \cite{mislove2007measurement}). 
This type of data is mainly a variety of network topologies, such as Kronecker \cite{leskovec2010kronecker}, Block Two-Level Erdos-Renyi \cite{PhysRevE.85.056109}, and Uncorrelated Scale-Free Networks \cite{PhysRevE.71.027103}. 
They are indispensable because researchers can control the complexity of the propagation scenarios \cite{RevModPhys.87.925}. 
For example, topics can be set as irrelevant in synthetic networks in terms of topic-related diffusion models, which reduces the modeling difficulty. 
The \emph{real-world data} are collected from social platforms (\textit{e.g.}, online social networks, email networks, academic collaboration networks). 
They contain more detailed information such as comments and user-related data, which provides rich contextual semantics for diffusion modeling. 
With the thriving of machine learning, this type of data has become the preferred one. 
Besides crawled by selves, these data also can be downloaded from dataset websites \cite{snapnets, aminer, Vlado_data}. 
Comparison of real-world and synthetic data is shown in Fig. \ref{FIG:RealAndSynthetic}. 
Some synthetic data are similar to real data in terms of statistical laws.

\subsubsection{Diffusion Mechanism Analysis} 
After collecting and processing data, the second step is to analyze the mechanism of diffusion processes, guiding the diffusion modeling and feature extraction. 
The mechanism can be analyzed from three aspects in accordance with the individual interaction scenario shown in the right part of the Fig. \ref{FIG:DiffusionScene}: \textbf{social vector-scope}, \textbf{user (behavior)-scope}, and \textbf{information-scope}. 
The \emph{social vector-scope} analysis focuses on studying the impact of underlying network structure on information diffusion. 
\textit{First}, users in different locations play divergent roles in the propagation process \cite{Bakshy2012theRoleOf,lou2013mining, yang2015rain}. 
\textit{Second}, different platforms may lead to various diffusion dynamics. 
For example, the same story spread faster on Digg, but spread more widely on Twitter, and the network of Digg is denser \cite{Lerman2010InfoContagion}. 
\textit{Third}, the research shows that social actions will promote the evolution of the social networks \cite{li2016exploiting, susarla_social_2012}, and social network evolution does impact user's social behaviors \cite{Althoff2017OnlineActions}. 
The \textbf{user (behavior)-scope} analysis concentrates on predicting the individual behavior (\textit{e.g.} like, retweet, and follow) in terms of specific information. 
What decision the user will make, depends on the match of the user cognition and information semantics. 
Although the brain's decision-making process is extremely complicated \cite{smith2008cognitive, Anderson2007-ANDHCT}, it is definite that analyzing user decision-making processes can promote the accuracy of information diffusion modeling. 
The \textbf{information-scope} analysis is commonly carried out from two levels: statistical characteristics (\textit{e.g.} the number of hashtags/pictures \cite{suh2010want}) and semantics (\textit{e.g.}, topic distribution, sentiment, emotion. ) 
Generally, researchers usually analyze information content and user behavior together. 

As shown in the right part of Fig. \ref{FIG:DiffusionScene}, the above three components affect the diffusion process and shape each other. 
First, users perceive the social information and others' attitudes from social vectors (The ``Receiving Information" and ``Peer Influence'' processes). 
Social vectors determine which information the user can receive, and whose the user can communicate. 
Second, they decide which kind of social action to take (The ``Decision'' process).  
Third, social actions can drive information propagation and social network evolution (The ``Information Propagation'' and ``Network Evolution'' processes). 
Generally, researchers should consider information, users, and network structures as a unification.

\subsubsection{Diffusion Modeling} 

From the perspective of modeling techniques, researchers adopt \textbf{time-series} and \textbf{data-driven} approaches. 
In the early stage of modeling researches, \textbf{time-series} approaches are meanstreams. 
Their modeling concepts are using explicit mathematical expressions to model the dynamics of information diffusion over time. 
\emph{Fitting}, \emph{simulation}, and \emph{likelihood maximization} are the three primary research directions. 
\emph{Fitting} models attempt to fit the curve of the diffusion volume over time. 
\emph{Simulation} ones try to imitate the information diffusion process based on social network topology.
Compared with the above two methods, \emph{likelihood maximization} ones construct a likelihood function based on the diffusion history and tries to infer the underlying diffusion networks. 
However, an inevitable defect of these time-series approaches is that the intrinsic diffusion dynamics are difficult to be expressed with formulas. 
Therefore, many researchers have shifted their focuses to data-driven approaches.  
\textbf{Data-driven} ones do not need to give specific expressions and use machine learning algorithms to get the prediction result automatically. 
More intrinsically, data-driven approaches show how to learn diffusion models, rather than give mathematical diffusion models explicitly. 
Researchers can manually extract diffusion features, or directly learn features from raw diffusion data in ``end-to-end'' approaches. 

To align with various modeling requirements, the outputs of information diffusion models include three levels of granularity: volume, individual, and the propagation relationship between two individuals, as shown in Fig. \ref{FIG:Methodology} 
Volume models only capture changes in the diffusion volumes over time, and do not care about who is infected. 
Individual adoption models attempt to identify every infected user. 
Relationship inference models aim at inferring the information diffusion track among those infected users. 
The above three levels of granularity models will be discussed in Section \ref{sec:TaxonomyDiffusionModels}. 

\subsubsection{Feature Extraction}
\label{sec:RMP:FeatureExtraction} 

The fourth critical step of the information diffusion research is to extract representative and comprehensive features from the collected data for the established model. 
Commonly used features can be divided into four categories: structural, temporal, content, and user-related. 
Researchers found that not all features play an equal role in diffusion prediction. 
Chen \textit{et al.} \cite{cheng2014can} verified that when structural features and temporal features are used in combination, the prediction accuracy can reach more than 70\%. 
Besides, experiments performed in \cite{barbieri2013topic} show that the prediction accuracy of the topic-aware model is 28\% higher than that of the topic-blind model. 
Some representative features are summarized in Appendix \ref{sec:Appendix}. 

\textbf{Structural:}  Network topology is the medium of information diffusion, which affects the trajectory and range of information diffusion. 
Existing structural features contain two types of categories: network structure, and cascade structure. 
\begin{itemize}
  \item Network structure features describe characteristics of social vectors (\textit{i.e.,} network topology), such as degree distribution \cite{hong2011predicting}, density \cite{ma2012will}, and the fraction of users forming triangles \cite{ma2013predicting}. 
  \item Cascade structure features refer to the characteristics of diffusion dynamics in social networks, involving retweet tree depth \cite{cheng2014can}, number of diffusion hops \cite{yang2010predicting}, retweet ratio \cite{tsur2012s}, and so on.
\end{itemize}

\textbf{Temporal:} Temporal features indicate the speed of information diffusion in social networks. 
They are mainly composed of Sequential features and statistical features. 
\begin{itemize}
  \item Sequential features describe numerical changes of some indicators over time during the diffusion process. 
  These indicators can be the number of reposts/comments/views \cite{yang2010predicting,cheng2014can,wu2018adversarial}, time elapsed between the current and previous post \cite{hong2011predicting, cheng2014can}, and so on.
  \item Statistical features are secondary processing of sequential features, aiming to find some potential diffusion laws. 
  Scholars usually analyze time-series features from macroscopic perspectives, such as the similarity between two time-series \cite{ahmed2013peek}, the shape of time series \cite{kong2014predicting,figueiredo2016trendlearner}, and so on.
\end{itemize}

\textbf{Content:} Information content can naturally affect the information diffusion process. 
At present, scholars extract content features from both statistical and semantic aspects. 
\begin{itemize}
  \item Statistical features mainly related to message forms, such as the number of hashtags/URLs \cite{suh2010want}, language \cite{cheng2014can}, and terms frequency \cite{naveed2011bad}. 
  \item Semantic features refer to the deep semantics of a message, including topics \cite{hong2011predicting}, sentiment \cite{ferrara2015quantifying}, emotions \cite{pfitzner2012emotional}, self-disclosure \cite{yuan2016will}, and so on. Most of these features need to be extracted using Natural Language Processing (NLP) and multi-modal techniques. 
\end{itemize}

\textbf{User-related:} 
Currently, user-related features are mainly composed of two parts: static features and dynamic features. 
\begin{itemize}
  \item Static features also can be called user profiles, mainly refer to user's inherent attributes, such as age, gender, country, and education background. 
  \item Dynamic features are mainly extracted from user interaction history, including preference \cite{zhang2013social}, participation engagement \cite{suh2010want}, and the probability of his tweet to be retweeted \cite{hong2011predicting}.
\end{itemize}

\subsubsection{Model Evaluation} 
As shown in Fig. \ref{FIG:Methodology}, the diffusion model is either regression or classification model. 
Commonly used evaluation indicators are also applicable here. 
Due to space limitations, we will not describe it in detail here.

%---------------------------------------------------------------
% SECTION 3 TAXONOMY OF DIFFUSION MODELS
%---------------------------------------------------------------

\section{Taxonomy of Elementary Diffusion Techniques} 
\label{sec:TaxonomyDiffusionModels}

\begin{figure}[h]
  \centering
  \includegraphics[width=5in]{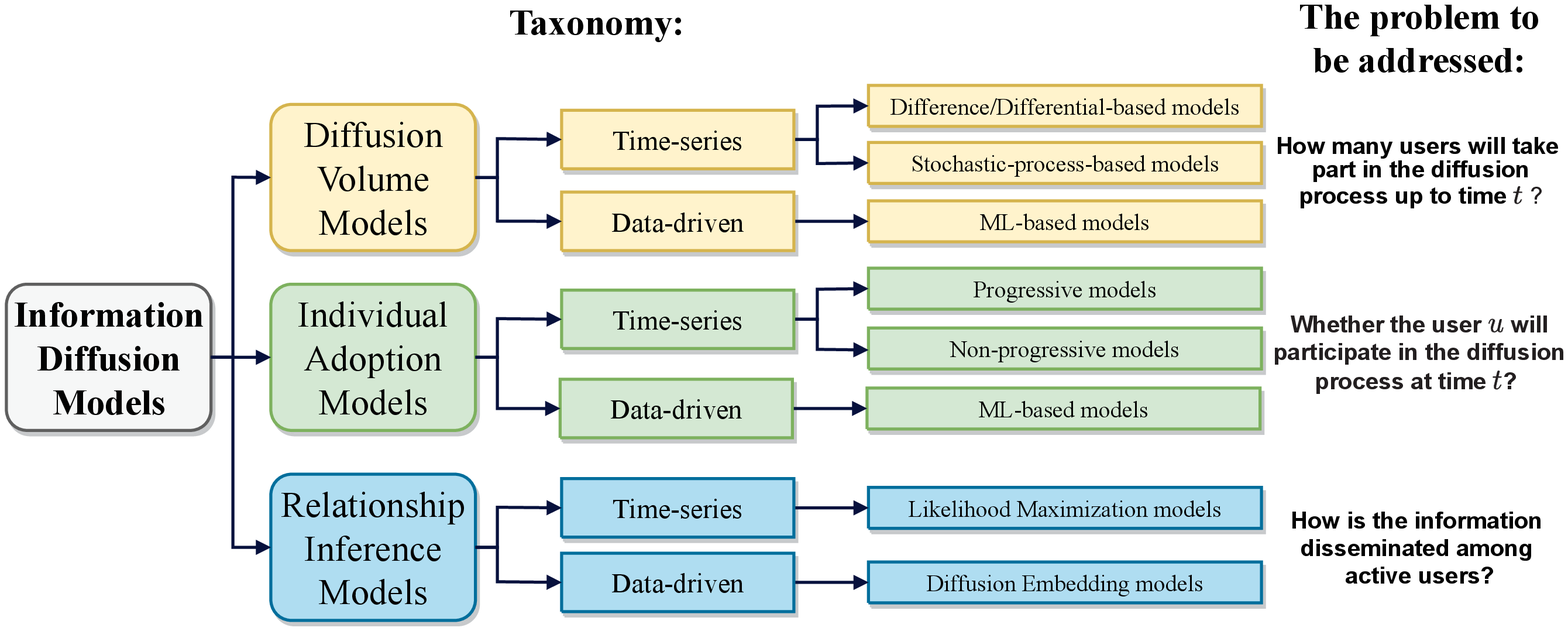}
  \caption{Taxonomy of current information diffusion models.}
  \label{FIG:Taxonmy}
\end{figure}

In this section, we thoroughly review techniques for information diffusion modeling and discuss their pros and cons. 
The existing diffusion models are essentially classified into three categories: diffusion volume models, individual adoption models, and relationship inference models. 
Diffusion volume models aim at predicting the overall diffusion volume, ignoring whether a specific user will receive the message.  
Individual adoption models attempt to identify future active users. 
Relationship inference models hope to clarify the propagation relationship between two active users. 
These three types of models can be established via time-series and data-driven approaches. 
The taxonomy is shown in Fig. \ref{FIG:Taxonmy}. 

\subsection{Diffusion Volume Models}
\label{sec:TDM:DiffusionVolumeModels} 

Existing diffusion volume models fall into three categories: difference/differential-based, stochastic-process-based, and ML-based models. 
These three types of models will be discussed as follows.

\subsubsection{Difference/differential-based Models}
\label{sec:TDM:DVM:Difference}

These models are the earliest diffusion models borrowed from Epidemics \cite{RevModPhys.87.925}. 
They divide the population into several compartments (or classes, status) such as S (Susceptible), I (Infected), R (Recovered), according to the different user statuses in the diffusion process. 
Next, difference or differential equations are used to describe the evolution of user proportions among each compartment. 

\textbf{Assumptions:} Difference/differential-based models are premised on that the total population is known, and each user connects with others effectively at each time unit. 

\textbf{Methods:}
The universal idea is to divide users into different compartments, and then build difference/differential equation models based on the user’s social connectivities to describe the evolution of each compartment.  
Common transition rules of compartments include Susceptible-Infected, Susceptible - Infected - Susceptible \cite{SIS_SIR_newman2003structure}, Susceptible - Infected - Recovered \cite{SIS_SIR_newman2003structure}, Susceptible-Exposed-Infected-Recovered \cite{LI1995155}, and Susceptible-Exposed-Infected-Removed \cite{XWYZ_huo2017dynamical}, etc. 
There are corresponding transition probabilities between different compartments. 
For example, SIR can be sketched in Algorithm. \ref{ALG:WorkflowOfSIS}. 
All users are divided into different $K$ groups according to their social connectivities (\textit{e.g.}, node degree \cite{PhysRevLett.86.3200, Bogua2003}, friendship hops \cite{PDE_wang2012diffusive}, and connectivity \cite{xu2015epidemic}). 
$p_k$ and $q_k$ are infectious rate and recovery rate of $k$-th group, respectively. 
$\varTheta_k(t)$ denotes the probability that any given edge is connected to an infected user in $k$-th group. 
Generally, the form of $\varTheta_k(t)$ is related to the network topology, such as degree distribution \cite{PhysRevLett.86.3200, Bogua2003, xu2015epidemic}, and distribution of friendship hops \cite{PDE_wang2012diffusive}. 

\begin{algorithm}[h]
  \caption{Basic Ideas of the SIR}
  \label{ALG:WorkflowOfSIS}
  \LinesNumbered
  \KwIn{$P(k), k\in\left[1,K\right]$: Distribution of node connectivity, $K$ is the maximum level of connectivity; $p_k$: Infectious rate; $q_k$: Recovery rate; $T$: Deadline}
  \KwOut{$S_k(T), I_k(T), R_k(t)$: Proportions of different class of users}

  $S_k(0), I_k(0),R_k(0) \Leftarrow$ Initialize user distribution with $P(k)$\;
  $t = 0$\;
  \While{$t<T$}{ 
    // Update user proportions\;
    $S_k(t+1) = -p_k S_k(t) \varTheta_k(t)$ \;
    $I_k(t+1) = p_k S_k(t) \varTheta_k(t) - q_k R_k(t)$\;
    $R_k(t+1) = q_k R_k(t)$\;
    $t=t+1$\;
  }

\end{algorithm}

Classical Epidemic models assume that all users are homogeneous. 
Therefore, the transition probability of any user in the same transition schema (\textit{i.e.}, the source compartment and target compartment are the same respectively) is identical. 
For example, as the elementary Epidemic model, SIR \cite{SIS_SIR_newman2003structure}, the number of newly infected people at time \(t\), \(\Delta I(t)\), equals the infection rate multiply the number of connections among susceptible and infected users. 
In other words, they are special cases of universal difference/differential-based models, where $K=1$ and $\varTheta_k(t)$ is the proportion/number of infected users. 

Besides these elementary difference/differential expressions, researchers develop many extended models, as shown in Table \ref{TAB:SummaryOfDiff}. 
Yang \textit{et al.} \cite{SIIS_yang2017bi} considered a generalized infection probability in a multi-virus competing scenario because linear infection rates may cause the diffusion volume overestimation. 
Matsubara \textit{et al.} \cite{SPIKEM_matsubara2012rise} introduces dynamic infection rate and periodicity to describe the six types of ``rise and fall'' patterns in social networks \cite{yang2011patterns}. 
They employed the Sine function to model the periodicity. 
Stai \textit{et al.} \cite{stai2018temporal} introduced external exposure which makes susceptible users become infected from other platforms (\textit{e.g.,} TV news), accelerating information diffusion process.

Notably, user payoffs models divide users into different groups according to their social strategies. 
They assume user social strategies are determined by payoffs (\textit{e.g.}, how many fans will be attracted if they forward the message?). The payoff can be computed based on the payoff matrix associated with social links. 
And then, game theory can be employed to model interactions among different groups.  
Jiang \textit{et al.} \cite{jiang2014graphical, jiang2014evolutionary} believed that the payoff matrix is symmetrical because they held the view that if a user adopting a forwarding strategy meets a user who takes a non-forwarding strategy, each of them will get the same payoff. 
Also, they modeled the information diffusion process using the evolutionary game-theoretic framework. 
Users with new information are regarded as mutants, and the diffusion process can be considered as mutant dissemination in the network. Thus the model can predict the final stable state of the network.
Cao \textit{et al.} \cite{cao2016evolutionary} classified user neighbors into two categories: known types, and unknown types. 
For neighbor users \(u\) and \(v\), if \(v\) does not know \(u\), payoff matrices of \(v\) have nothing to do with \(u\), and vice versa.

\begin{table}[!h]
  \scriptsize
  \caption{\label{TAB:SummaryOfDiff}Summary of Difference/differential-based Models}
  \newcommand{\tabincell}[2]{\begin{tabular}{@{}#1@{}}#2\end{tabular}}
  \begin{tabular}{ p{1.5cm} p{1cm}  p{4cm} p{6.2cm}}
    \toprule
    \tabincell{c}{\textbf{Extentions}} &\tabincell{c}{\textbf{Articles}}    & \tabincell{c}{\textbf{Approaches}} & \tabincell{c}{\textbf{Descriptions}}\\ \midrule

    \tabincell{l}{Dynamic\\Infection Rate} & \tabincell{c}{\cite{SPIKEM_matsubara2012rise,SIIS_yang2017bi}} & \tabincell{c}{$p_u \Longrightarrow f_u(t)$}   & \tabincell{l}{Make the infection rate decays in an exponential distribution \\ \cite{SPIKEM_matsubara2012rise}, or turn the linear transition rate into a twice continuously \\differentible function \cite{SIIS_yang2017bi}}  \\ \midrule

    \tabincell{c}{Periodicity} & \tabincell{c}{\cite{SPIKEM_matsubara2012rise}} &\tabincell{c}{$f^{\text{period}}=1-\frac{1}{2}P_a(\sin (\cdot))$}     & \tabincell{l}{  Design a  function of periodicity based on the Sine function. \\$P_a$ is strength of periodicity. }  \\ \midrule

    \tabincell{c}{External \\Exposure}    &\tabincell{c}{\cite{stai2018temporal}} & \tabincell{c}{$\frac{dI(t)}{dt} = I(t) S(t) p_1(t)$\\$ + S(t)+p_2(t)$} & \tabincell{l}{Import time-varying external exposure probability $p_2(t)$.}  \\ \midrule

    \tabincell{c}{User\\Payoffs}  &\tabincell{c}{\cite{jiang2014graphical,jiang2014evolutionary,cao2016evolutionary}} & \tabincell{c}{$\frac{dI_i(t)}{dt}= \varphi_i(t)-\varphi(t)$}& \tabincell{l}{Divide users into different groups based on their social \\ strategies. And then, use game theory to model the evolution \\dynamics on these groups. $\varphi_i$ and $\varphi$ are the average payoffs of \\group $i$ and total population, respectively. }  \\ \bottomrule

  \end{tabular}
\end{table}

\textbf{Pros and Cons:} 
Difference/differential-based models are straightforward, efficient, extendable, and conforming to some real-world physic laws \cite{RevModPhys.87.925}. 
They first define the user state transition pattern in diffusion modeling. 
Especially, since they do not involve specific individuals, they perform well in macroscopic diffusion modeling. 
For example, PDE \cite{PDE_wang2012diffusive} reflects that the density of infected users decreases as the friendship distance increase. 
SpikeM \cite{SPIKEM_matsubara2012rise} captures the power-law fall pattern and periodicities, and avoids the divergence to infinity. 
On the one hand, this abbreviation reduces the modeling complexity and makes these models indispensable. 
Taking SIS as an example, the epidemic threshold of homogeneous networks is the inverse of the average node degree \cite{diekmann2000mathematical}, while the threshold is absent in Scale-Free networks \cite{PhysRevLett.86.3200}. 
On the other hand, they do not consider the heterogeneity of user activity \cite{PhysRevLett.103.038702},  which makes them difficult to use for individual-level tasks. 
In addition, Iribarren \textit{et al.} \cite{PhysRevLett.103.038702} considered that oversimplified models can not accurately model the information diffusion process because they are difficult to reflect user heterogeneity.

\subsubsection{Stochastic-process-based Models}
\label{sec:TDM:DVM:Stochastic}
Compared with difference/differential-based models, stochastic-process-based ones estimate the diffusion contribution made by each reshare event or each user.

\textbf{Assumptions:}
These models assume that the intensity of social events (\textit{e.g.}, message arrival, or user posting/resharing message) occurring at each time \(t\) can be characterized by stochastic processes. 
Suppose there are \(I(t)\) events that occurred as of time \(t\) (\textit{i.e.}, there are $I(t)$ users reshare the item). 
The intensity is the conditional probability that an event occurs in the time interval \([t, t+\Delta t]\) on the condition that it does not happen until time \(t\). 
Moreover, the diffusion volume can be calculated by integrating the intensity at each time \(t\). 
Let the random variable, \(\tau\), denote the waiting time until an expected event occurs. 
The intensity can be expressed as Eq. \ref{EQU:Intensity} based on Survival analysis \cite{kleinbaum2010survival}. 

\begin{scriptsize}
  \begin{equation}
    \lambda(t)= \lim\limits_{\Delta t \to 0} \frac{P(t < \tau < t+\Delta t | \tau \geq t )}{dt}
    \label{EQU:Intensity}
  \end{equation}
\end{scriptsize}

\begin{table}[h]
  \newcommand{\tabincell}[2]{\begin{tabular}{@{}#1@{}}#2\end{tabular}}
  \caption{\label{TAB:SummaryOfStochastic}Summary of Stochastic-process-based Models}
  \scalebox{0.9}{
    \renewcommand\arraystretch{1}
    \begin{tabular}{p{1.8cm} p{1.8cm} p{3cm} p{3cm} p{4cm}}
      \toprule
      \tabincell{c}{\textbf{Mechanisms}} &  \tabincell{c}{\textbf{Models}} & \tabincell{c}{\textbf{Simplified}\\\textbf{Expressions}}  & \tabincell{c}{\textbf{Articles}} & \tabincell{l}{\textbf{Parameters}} \\ \midrule

      \multirow{5}{*}{\tabincell{c}{Decay}} & \tabincell{c}{Power-law} & \tabincell{c}{$(\frac{t-t_i}{c})^{-(1+\theta)}$} & \tabincell{c}{\cite{Gao:2015:MPR:2684822.2685303,SEISMIC_zhao2015seismic,kobayashi2016tideh,mishra2016feature}\\\cite{DBLP:conf/www/RizoiuXSCYH17,10.1145/3289600.3291601}}& \multirow{5}{*}{\tabincell{l}{$c$: bounded term of minimum \\time difference \\ $\theta, \theta^{'}$: model parameters  \\ $\Phi$: CDF of the Normal \\distribution}} \\ \cmidrule{2-4}
      & \tabincell{c}{Exponential} & \tabincell{c}{$e^{-\theta(t-t_i)}$} & \tabincell{c}{\cite{bao2015modeling,Bao:2016:MPP:2983323.2983868}} & \\ \cmidrule{2-4}
      & \tabincell{c}{Rayleigh} & \tabincell{c}{$e^{-\theta\frac{(t-t_i)^2}{2}}$} & \tabincell{c}{\cite{10.1093/aje/kwh255}} & \\ \cmidrule{2-4}
      & \tabincell{c}{Log-Normal} & \tabincell{c}{$1-\Phi(\frac{\ln(t-t_i)}{\sigma})$} & \tabincell{c}{\cite{shen2014modeling}} & \\ \cmidrule{2-4}
      & \tabincell{c}{Weibull} & \tabincell{c}{$e^{-(\frac{t-t_i}{\theta})^{\theta^{'}}}$} & \tabincell{c}{\cite{yu2015micro}} & \\ \midrule

      \multirow{2}{*}{\tabincell{c}{Exciting}} & \tabincell{c}{Poisson} & \tabincell{c}{\\$\sum_{j}^{j=k}e^{-\beta j}$\\\quad} & \tabincell{c}{\cite{shen2014modeling, Gao:2015:MPR:2684822.2685303}}& \multirow{2}{*}{\tabincell{l}{$j,k$: $j,k$-th retweet events \\$\phi(t-t_j)$: decay function \\ $\alpha_j$: branch factor (\textit{i.e., triggering} \\\textit{strength of each forwarding}) \\ $\beta$:model parameter. If $beta=0$, \\it is Linear exciting function. }} \\ \cmidrule{2-4}
      & \tabincell{c}{Hawkes} & \tabincell{c}{\\$\sum_{j}^{j=k}\beta_j \phi(t-t_j)$\\\quad} & \tabincell{c}{\cite{bao2015modeling,Bao:2016:MPP:2983323.2983868,SEISMIC_zhao2015seismic,kobayashi2016tideh,mishra2016feature}\\\cite{DBLP:conf/www/RizoiuXSCYH17,10.1145/3289600.3291601}} & \\ \midrule

      \tabincell{c}{Periodicity} & \tabincell{c}{Sine} & \tabincell{c}{$\beta \sin (\frac{2\pi}{\phi}(t+s))$} & \tabincell{c}{\cite{Bao:2016:MPP:2983323.2983868, kobayashi2016tideh}}& \tabincell{l}{$\beta$: strength of periodicity \\ $\phi$: phase of periodicity \\ $s$: phase shift of periodicity} \\ \midrule

      \multirow{2}{*}{\tabincell{c}{External\\Influence}} & \tabincell{c}{Sine} & \tabincell{c}{\\$\beta \sin (\frac{2\pi}{\phi}(t+s))$\\\quad} & \tabincell{c}{\cite{Bao:2016:MPP:2983323.2983868}}& \multirow{2}{*}{\tabincell{l}{$\alpha,\beta$: strength of periodicity \\ $\phi$: phase of periodicity \\ $s$: phase shift of periodicity \\ $\mathbb{I}$:impulse function \\ $\bar{s}(t)$:number of observable \\external influence}} \\ \cmidrule{2-4}
      & \tabincell{c}{Impluse\\function} & \tabincell{c}{\\$\alpha \mathbb{I}\left[t=0\right] + \beta \mathbb{I} \left[t>0\right]$\\$ + \bar{s}\left[t\right]$\\\quad} & \tabincell{c}{\cite{DBLP:conf/www/RizoiuXSCYH17}} & \\ \bottomrule

    \end{tabular}
    }
\end{table}

\textbf{Methods:} 
Every model can be decomposed into three parts: \textit{basic intensity, decay, and exciting}. 
The \textit{basic intensity} indicates the most basic diffusion intensity, which can be implicitly represented as a part of the model. 
It can reflect some diffusion phenomena, such as intrinsic infectiousness, periodicity, and external influence. 
The \textit{decay} function indicates that the item's influence decrease with time. 
It can also be understood as human reaction time \cite{SEISMIC_zhao2015seismic}. 
The \textit{exciting} function enables each previous forwarding to have a continuous impact on the following interactive behaviors, shedding light on the ``richer-and-richer'' phenomena \cite{crane2008robust}. 
We summarized the motivations of these models and organized them into Table \ref{TAB:SummaryOfStochastic}.

For decay functions, the power-law model is often used to indicate the long-tails in social networks \cite{rodriguez2011uncovering}. 
Other models will perform better in some datasets, such as Exponential in microblog \cite{bao2015modeling,Bao:2016:MPP:2983323.2983868}, Log-Normal in citation network \cite{shen2014modeling}, and Rayleigh in epidemiology \cite{10.1093/aje/kwh255}. 
Especially, Yu \textit{et al. } \cite{yu2015micro} found that the Weibull distribution can preserve the minor and early-stage dominance better.

Commonly used \textbf{exciting mechanisms} are Poisson and Hawkes, as shown in Eq. \ref{EQU:Possion} and \ref{EQU:Hawkes}, respectively. 
The exciting component of Hawkes-based models implicitly represented by $\sum$. 
\begin{scriptsize}
  \begin{equation}
    \lambda(t) = \lambda_{base} f_{decay}(\cdot) g_{excit}(\cdot)
    \label{EQU:Possion}
  \end{equation}
\end{scriptsize}
\begin{scriptsize}
  \begin{equation}
    \lambda(t)=\lambda_{base} + \sum_{i=1}^{N(t)}f_{decay}(\cdot)
    \label{EQU:Hawkes}
  \end{equation}
\end{scriptsize}

For \textbf{Poisson-process-based} models, Shen \textit{et al.} \cite{shen2014modeling} assumed that the excitement of each event is identical, and adopted the number of attentions the message received as the Linear exciting function. 
Gao \textit{et al.} \cite{Gao:2015:MPR:2684822.2685303} applied exponential to capture the decay effective of previous retweet. 
However, these methods stay at the level of activation index, instead of involving specific activation time. 
\textbf{Hawkes-based} models provide the self-exciting mechanism, and the triggering strength can be calculated based on the activation time interval.    
SEHP (Self-Excited Hawkes Process) \cite{bao2015modeling} uses the exponential expression to compute the base intensity and decay effects. Moreover, ISEHP (Influence-based Self-Excited Hawkes Process) \cite{Bao:2016:MPP:2983323.2983868} incorporates the periodicity using \textit{sin} function.
Zhao \textit{et al.} \cite{SEISMIC_zhao2015seismic} models the time-varying infectiousness and the arriving time of each post are modeled by two Hawkes processes, respectively.
It is able to identify at each time point whether the cascade will be "explosive" or "tractable" according to whether its infectiousness is above or below a critical threshold. 
Kobayashi \textit{et al.} \cite{kobayashi2016tideh} thought that the final cascade size can be achieved by integrating the future retweet rate \(\hat{\lambda}(t)\) which is calculated based on the average number of followers and the infectious rate.
Mishra \textit{et al.} \cite{mishra2016feature} used a predictive layer on top of the MHP (marked Hawkes process) to make predictions with the combination of user social influence, social memory, and tweet quality. 
Marian \textit{et al.} \cite{DBLP:conf/www/RizoiuXSCYH17} introduced the external influence modeled by impulse function, which allows their model to fit complex situations. 
They also proved that the intensity $\lambda(t)$ is identical to the infection probability $p$ after marginalizing out recovery events, and they proposed the HawksN model with finite population \cite{10.1145/3289600.3291601}. 

Generally, the modeling process of stochastic-process-based models is roughly divided into the following basic steps, as shown in Algorithm \ref{ALG:WorkflowOfStochasticProcess}. 
\textit{First}, model the distribution of influence decay $f_{decay}(t)$ with reference to decay functions. 
\textit{Second}, devise the intensity $\lambda(t)$ based on Survival theory and $f_{decay}(t)$. 
In this step, we can add some other features to the model. 
\textit{Third}, estimate model parameters via optimization methods such as E-M (Expectation-Maximization). 
\textit{Fourth}, compute the influence of each social event by integrating the intensity $\lambda(t)$ from start to end. 
\textit{Last}, the final diffusion volume can be calculated with the function $\Phi(\cdot)$.  (The last step is optional according to model requirements.)

\begin{algorithm}[h]
  \caption{Basic Ideas of Stochastic-process-based Models}
  \label{ALG:WorkflowOfStochasticProcess}
  \LinesNumbered
  \KwIn{\(c = \left\{ (u_1,t_1); (u_2, t_2);\ldots \right\}\): A piece of cascade; $T$: Last time $t$ of observable cascade}
  \KwOut{\(I^c(T^{'})\): Final cascade size of the cascade $c$}

  $f_{decay}(t) \Leftarrow $ Model the distribution of influence decay \;
  $\lambda(t) \Leftarrow $ Devise the intensity of social events\;

  $i=0$\;
  \For{$i<|c|-1$}{
    $\delta_i = t_{i+1}-t_i$\;
    $\eta_i = t_{T}-t_i$\;
    \vdots
  }
  
  $\lambda(t) \Leftarrow \{\{\delta_i \},\{\eta_i \}, \ldots \}$ // Parameter estimation\;

  \For{$i \in \{j | t_j < T \ \textbf{and} \ u_j \in c\}$}{
    $R(i) = \int_{t_0}^{T} \lambda(t)dt $ // Compute the influence of each social event\;
  }

  $I^c(T^{'}) = \Phi(\{R(i)\})$ // Compute the final cascade size prediction\;

\end{algorithm}

\textbf{Pros and Cons:} Stochastic-process-based models are straightforward and exhibit some natural diffusion phenomenons. 
These models show multiple decay effects which describe that the influence of one information decreases over time, including Power-Law \cite{Gao:2015:MPR:2684822.2685303, mishra2016feature, SEISMIC_zhao2015seismic}, Exponential \cite{bao2015modeling, Bao:2016:MPP:2983323.2983868}, Log-Normal \cite{shen2014modeling}, and Oscillation \cite{kobayashi2016tideh}. 
Furthermore, exciting functions reflect that past events will promote the event occurring in the future. 
The introduction of time factors makes model performance improve around 30\% \cite{kobayashi2016tideh}. 
Moreover, whether the content will be popular is related to exogenous sensitivity, and endogenous response \cite{crane2008robust, DBLP:conf/www/RizoiuXSCYH17}. 

Notably, if the intensity involves every specific user, the next infected user can be predicted with the help of likelihood maximization inference \cite{Iwata:2013:DLI:2487575.2487624}. 
Nevertheless, there are few such models, so that we will not be separately discussed. 
Besides, these models require sophisticated parameter estimation. 
Although complex parameter estimation is needed, the time complexity of most models is still close to linear \cite{SEISMIC_zhao2015seismic, yu2015micro}.

\begin{table}[h]
  \newcommand{\tabincell}[2]{\begin{tabular}{@{}#1@{}}#2\end{tabular}}
  \caption{\label{TAB:SummaryOfDeepVolume}Summary of Deep Learning (Volume) Models}
  \scalebox{0.8}{
    \renewcommand\arraystretch{1}
    \begin{tabular}{p{1.8cm} p{3.7cm} p{5.5cm} p{3cm}}
      \toprule
      \tabincell{c}{\textbf{Scopes}} &  \tabincell{c}{\textbf{Key Components}} & \tabincell{c}{\textbf{Contributions}}  & \tabincell{c}{\textbf{Articles}} \\ \midrule

      \tabincell{c}{Temporal} & \tabincell{c}{LSTMIC\\(LSTM+ Pooling)} & \tabincell{l}{LSTM: Embed the sequential features. \\Pooling: Extract high-level sequential \\patterns.} & KAIS 2018 \cite{DBLP:journals/kais/GouSDWLC18} \\ \midrule

      \multirow{2}{*}{\tabincell{c}{Content \& \\User-related}} & \tabincell{c}{UHAN\\(Embedding + \\ Hierarchical Attention)} & \tabincell{l}{Embedding: Embed visual, textual, \\and publisher (user) by VGGNet, \\LSTM, and Matrix Embedding\\, respectively. \\ Hierarchical Attention: Compute \\textual and visual embedding as \\well as their importance by intra \\and inter attention, respectively. } & WWW 2018 \cite{DBLP:conf/www/ZhangWWZ18} \\ \cmidrule{2-4}
      & \tabincell{c}{MOOD\\(Hierarchical Attention \\Embedding \\+ Tensor Factorization)} & \tabincell{l}{Hierarchical Attention Embedding:\\ Obtain the word-level \\representation for diffusion \\context (\textit{e.g.,} organizor, location). \\Tensor Factorization: Compute \\the joint influence of different \\components of diffusion context.} & DASFAA 2018 \cite{DBLP:conf/dasfaa/WangZW18} \\ \midrule

      \multirow{4}{*}{\tabincell{c}{Structural \\\& Temporal}} &  \tabincell{c}{DeepCas \\(Random Walk + \\GRU + Attention)} & \tabincell{l}{Random Walk: Sample multiple \\sequences from a cascade graph\\GRU: Encode sequence to a \\(hidden) vector  \\ Attention: Aggregate sequence \\hidden representations to \\represent cascade graph}  & WWW 2018 \cite{10.1145/3038912.3052643}\\ \cmidrule{2-4}

      &  \tabincell{c}{DeepHawkes\\(Embedding \\+ GRU \\+ Sum Pooling)} & \tabincell{l}{Embedding: Embed user\\ GRU: Encode cascade path. \\ Sum Pooling: Integrate decay effect}  & CIKM 2017 \cite{cao2017deephawkes}\\ \cmidrule{2-4}

      & \tabincell{c}{CasCN\\(GCN + LSTM)} & \tabincell{l}{GCN: Embed sub-cascade graph \\by CasLaplacian. \\LSTM: Model the temporal \\dependency of cascades.  } & ICDE 2019 \cite{DBLP:conf/icde/Chen0ZTZZ19} \\ \cmidrule{2-4}

      & \tabincell{c}{VaCas\\(Spectral Graph Wavelet\\ + Bi-GRU + VAE)} & \tabincell{l}{SGW: Embed cascade subgraph \\ GRU: Embed temporal information \\ VAE: Model diffusion uncertainty  } & INFOCOM 2020 \cite{DBLP:conf/infocom/0002XZTZ20} \\ \midrule

      \multirow{2}{*}{\tabincell{c}{Temporal \& \\Content}} & \tabincell{c}{DTCN\\(Embedding + LSTM \\+ Attention)} & \tabincell{l}{Embedding: Obtain the unified \\representation of Multi-modal \\social sequences.  \\LSTM: Learn sequential and temporal \\coherence of a single cascade. \\ Attention: Infer the correlation \\among multiple cascades.} & IJCAI 2017 \cite{DBLP:conf/ijcai/WuCZHLM17} \\ \cmidrule{2-4}

      & \tabincell{c}{DFTC\\(LSTM + attention CNN \\+ Hierarchical Attention \\Network + Embedding+ \\Temporal Attentive Fusion)} & \tabincell{l}{LSTM+attention CNN: Embed global \\trend and local fluctuation of \\temporal features, respectively. \\ HAN: Embed the content. \\Embedding: Embed meta-data.   \\ Attentive Fusion: Integrate multiple \\features.  } & AAAI 2019 \cite{DBLP:conf/aaai/LiaoXLHLL19} \\  \bottomrule

    \end{tabular}
    }
\end{table}

\subsubsection{ML-based (Volume) Models}
\label{sec:TDM:DVM:Feature}
Compared with the first two categories that mainly adopt time-series approaches, data-driven models use learning algorithms to model the diffusion process without any preset model expression. 
 They take a series of features possibly relevant to the diffusion process (see details in Section \ref{sec:RMP:FeatureExtraction}) as input and output the popularity prediction. 
  
\textbf{Assumption:} As far as we know, ML-based models relax almost all unrealistic assumptions in diffusion scenarios. 
However, their performance depends on whether the algorithm can effectively learn the propagation characteristics.

\textbf{Methods:} 
Data-driven models can be divided into two stages: \textbf{feature-based} and \textbf{deep learning}.  
\textbf{Feature-based} models mainly adopt native machine learning techniques, taking a series of elaborate hand-craft features as input and output cascade prediction results. 
These models include Generalized Linear Model \cite{suh2010want}, Linear Regression \cite{LIM_yang2010modeling}, Support Vector Machine \cite{ma2013on}, Decision Tree \cite{ma2012will, ma2013predicting},  Random Forest \cite{DBLP:conf/sbp/AlzahraniAKDT15, DBLP:conf/sdm/DingLLJ18}, Neural Networks \cite{DBLP:conf/mm/DingWW19}, and so on. 
Some literature \cite{cui2013cascading,cheng2014can,gao2019taxonomy} have shown that features affect prediction performance more than learning algorithms. 
They also found that temporal and structural features are critical predictors of cascade size. 
In addition, many researchers have considered features of early diffusion, especially some of the characteristics of the original reposters, which also have great help in the final popularity prediction.
Furthermore, in the initial stage, it is diffusion breadth, rather than depth, that has a more significant on large cascades \cite{cheng2014can}.

With the vigorous development of deep learning, researchers have gradually turned to end-to-end models, which can automatically learn sophisticated diffusion features. 
Representative models are introduced in Table \ref{TAB:SummaryOfDeepVolume}. 
Compared with feature-based models using native methods, \textbf{deep learning} models usually design new frameworks to utilize diffusion features better. 
The overall model framework is divided into four major steps: embedding data as vectors, using neural network models to learn diffusion features, aggregating features, and prediction. 
Generally, researchers adopt MLP (Multi Layer Perceptron) to predict diffusion volume. 
The key lies in the first three components of the framework. 

For temporal features, RNN (Recurrent Neural Network)-type technologies can model temporal dependency. 
Gou \textit{et al.} \cite{DBLP:journals/kais/GouSDWLC18} converted the retweeting time series to different perspectives (\textit{e.g.}, Frequency, Slope) with the help of time windows, then applied LSTM \cite{DBLP:journals/neco/HochreiterS97} to learn sequential temporal features, and used Pooling mechanism to extract high-level sequential features. 

For content and user-related features, publisher and multi-modal data can be unified representation via deep learning first, and then use the tensor operation technologies to learn correlation among these data. 
Zhang \textit{et al.} \cite{DBLP:conf/www/ZhangWWZ18} utilized VGGNet \cite{DBLP:journals/corr/SimonyanZ14a}, LSTM \cite{DBLP:journals/neco/HochreiterS97}, and matrix operation to obtain a unified representation of visual, textual, and personalized information. Then the hierarchical attention mechanism is adopted to learn the features and correlations of multi-modal information. 
Wang \textit{et al.} \cite{DBLP:conf/dasfaa/WangZW18} adopted the hierarchical attention mechanism to obtain the word-level representation of diffusion context (\textit{e.g.,} organizor, location), and then used tensor factorization to compute the joint influence of different components of diffusion context.

For structural and temporal features, researchers first embed the cascade graphs to learn structural features, and then utilize the RNN -like models to model sequential temporal features.  
Cheng \textit{et at.} \cite{10.1145/3038912.3052643} sampled individual cascade graph by Random Walk, and utilized GRU \cite{DBLP:journals/neco/HochreiterS97} to encode variable-length diffusion sequences as hidden representations. Finally, the attention mechanism is used to fuse these hidden representations for prediction. 
Cao \textit{et al.} \cite{cao2017deephawkes} implemented the decay mechanism via sum pooling, and Islam \textit{et al.} \cite{DBLP:conf/icdm/IslamMAPR18} utilized conditional intensity \cite{10.1145/2939672.2939875} to predict user activation time. 
Chen \textit{et al.} \cite{DBLP:conf/icde/Chen0ZTZZ19} developed CasLaplacian (a variety of spectral GCN \cite{DBLP:conf/nips/DefferrardBV16}) to embed cascade graphs, and utilized LSTM \cite{DBLP:journals/neco/HochreiterS97} to model the temporal dependency. 
Besides modeling structural and temporal dependency, Zhou \textit{et al.} \cite{DBLP:conf/infocom/0002XZTZ20} introduced VAE \cite{DBLP:journals/corr/KingmaW13} to model the diffusion uncertainty.

For temporal and content features, researchers first embed multi-modal social content and use RNN-like technologies to model temporal features. Then the attention mechanism is used to fuse these representations. 
Wu \textit{et al.} \cite{DBLP:conf/ijcai/WuCZHLM17} employed ResNet \cite{DBLP:conf/cvpr/HeZRS16} and FNN (Feed-forward Neural Network) to generate unified representations of multi-modal social data. And then, they adopted LSTM \cite{DBLP:journals/neco/HochreiterS97} to model neighboring and periodic temporal context. Finally, a multiple time-scale temporal attention method is proposed to implement cascade prediction. 
In addition, Liao \textit{et al.} \cite{DBLP:conf/aaai/LiaoXLHLL19} developed Attentive Convolution Neural Network (CNN) to capture the short-term fluctuation caused by external influence, and added a temporal decay factor to the loss function.

\textbf{Pros and Cons:} 
Compared with the above time-series models, data-driven ones focus on utilizing representative and comprehensive features rather than devising complex mathematical expressions. 
They enormously reduce the difficulty of diffusion mechanism analysis and diffusion modeling. 
In addition, they almost remove any preset assumptions, and they can get more than 95\% accuracy in multiple scenarios \cite{DBLP:conf/aaai/LiaoXLHLL19, DBLP:conf/sigir/ChenZ0TZZ19}.  
Especially, since there are no input restrictions, data-driven methods can discover useful features easier than time-series ones. 
However, their defects are obvious.
First, ML-based models require computational expensive feature engineering or learning representation. 
Second, their performance is highly dependent on the feature quality and quantity. Thus it is difficult to quantify the performance of these learning techniques with certainty. 
Third, existing works show that the prediction performance may be more sensitive as time goes on \cite{kong2014predicting}.

\begin{table}[!htbp]
  \caption{\label{TAB:SummaryOfDiffusionVolumeModels}Summary of Diffusion Volume Models}
  \newcommand{\tabincell}[2]{\begin{tabular}{@{}#1@{}}#2\end{tabular}}
  \centering
  \scalebox{0.9}{
  \begin{tabular}{p{1.8cm} p{2cm} p{3.5cm} p{2cm} p{2cm} p{2cm}}
    \toprule
    \multicolumn{2}{c}{}& \tabincell{c}{\textbf{Targets}} & \tabincell{c}{\textbf{Assumptions}}  &\tabincell{c}{\textbf{Explicit}\\\textbf{Expressions}}  &\tabincell{c}{\textbf{Key Factors}} \\
    \midrule 
    \multirow{2}{*}{\tabincell{c}{Time-series}} & \tabincell{c}{Difference/\\Differential\\-based} & \tabincell{c}{\(\Delta I(t)\) : Number of \\ newly infected users \\ at time \(t\)} & \tabincell{c}{Mean-Field \\Theory}   & \tabincell{c}{$\surd$} & \multirow{2}{*}{\tabincell{c}{Rationality of \\ hand-crafted \\ stochastic model}} \\ \cmidrule{2-5}

                                             & \tabincell{c}{Stochastic-\\process\\-based} & \tabincell{c}{\(\lambda_{u}(t)\) : Intensity of \\ diffusion event \\occurrence} & \tabincell{c}{Human \\Behavior \\Laws}    & \tabincell{c}{$\surd$} &  \\ \midrule

    \multirow{2}{*}{\tabincell{c}{ML-based}} & \tabincell{c}{Regression} & \tabincell{c}{\(I(t)\) : Number of total \\ active population \\ at time \(t\)} & \tabincell{c}{-}   &  \tabincell{c}{-} & \multirow{2}{*}{\tabincell{c}{Features extracted \\ from datasets}} \\ \cmidrule{2-5}

                                             & \tabincell{c}{Classification} & \tabincell{c}{\(I(t)>\tau\) : Whether the \\ diffusion volume at time \\\(t\) hits the threshold} & \tabincell{c}{-}    & \tabincell{c}{-} & \\ \bottomrule
  \end{tabular}
  }
\end{table}

\subsubsection{Analysis and Discussion}
The summary of diffusion volume models is shown in Table \ref{TAB:SummaryOfDiffusionVolumeModels}. 
Difference/differential-based models are straightforward, and their calculation efficiency is extremely high. 
It is noteworthy that their design philosophies (\textit{e.g.}, node state classification, and state transition rules) are widely adopted by other diffusion models. 
Compared with difference/differential-based models, which rely on the mean-field assumption, stochastic-process-based ones model the intensity of diffusion event occurrence. 
In other words, they model the influence of each user's forwarding on subsequent diffusion. 
Time-series models are interpretable, but it is challenging to devise appropriate mathematical expressions. 
ML-based approaches extract various features and use machine learning algorithms to predict the diffusion process. 
This characteristic allows them to mine more potential features. 
However, their predictive performance heavily depends on the quality of the extracted features, 
and whether their prediction performance is as stable as time-series models remain to be determined.

\subsection{Individual Adoption Models}
\label{sec:TDM:IndividualAdoptionModels} 
Individual-level models are essential because predicting the diffusion volume alone is not enough for specific user-involving scenarios such as viral marketing. 
They aim at predicting who will join the cascades, and all of them belong to classification task models.

\subsubsection{Progressive Models}
\label{sec:TDM:IAM:Progressive}
Similar to the SI model, progressive models divide all users into inactive and active groups. 
Active users will follow the rules specified by the model and infect their neighbors according to the social network topology. 

\textbf{Assumptions:} 
Progressive models believe that messages are propagated through social links from active users to inactive users, and active users will never be inactive again.

\begin{algorithm}[h]
  \caption{The Basic Idea of IC/LT Models}
  \label{ALG:WorkflowOfProgressive}
  \LinesNumbered
  \KwIn{$G=(V,E)$: Social topology; $\mathcal{I}(0)$: Seed users; $T$: Deadline}
  \KwOut{$\mathcal{I}(\textbf{T})$: User activation records from $t=0$ to $t=T$}

  $\mathcal{S} = \{u| u \in V \wedge  u \notin \mathcal{I}(0)\}$ // Initial susceptible user set 
  $t = 0$\;
  \While{$t<T$ and $|\mathcal{S}|>0$}{
    $t=t+1$\;
    \For{$u$ in $\mathcal{S}$}{
      \If{$\Phi(u,t)>0$}{ // Determine whether the user $u$ is activated 
        $\mathcal{I}(t) = \mathcal{I}(t) \cup \{u\}$\;
        $\mathcal{S} = \mathcal{S} - \{u\}$\;
      }

    }
  }

\end{algorithm}

\textbf{Methods:}
Independent Cascade (IC) \cite{IC_goldenberg2001talk} and Linear Threshold (LT) \cite{LT_granovetter1978threshold} are two early and elementary progressive models. 
They all assume that the network topology is static and propagations only occur at discrete-time synchronously. 
In addition, the user who is activated at time \(t\) only has one chance to infect all its neighbors at time \(t+1\).
Algorithm \ref{ALG:WorkflowOfProgressive} the sketch of the basic idea of IC/LT models. . 

Model difference is mainly reflected by the activate function $\Phi(u,t)$. 
In IC model, if the number of active neighbors of user \(v\) at time \(t-1\) is \(m\), the probability of user \(v\) will be activated at time \(t\) is 
\begin{scriptsize}
  \begin{equation}
    p_v(t)=1-\prod_{u\in\Delta\mathcal{I}(t-1) \land u\in N_{v}^{in}}(1-p_{uv})
    \label{EQU:IC}
  \end{equation}
\end{scriptsize}
The user \(v\) will be activated at time \(t\) with the probability $p_{v}(t)$. 
In the LT model, propagation probability can be thought of as the influence weight. 
Every user \(v\) is associated with a threshold \(\theta_{v}\). 
The user \(v\) will be activated if the Eq. \ref{EQU:LT} is satisfied. 
\begin{equation}
  \sum_{u\in \mathcal{N}_{I}(v)}{p_{u,v}} \geq \theta_{v}
  \label{EQU:LT}
\end{equation}

The IC/LT reflects some crucial characteristics of information diffusion. 
Typically, users \(u\) and \(v\) discuss the information \(I_{message}\) only once, and the user \(v\) may not agree with the user \(u\). 
If the user \(v\) adopts \(I_{message}\), he may forward it to his neighbors. 
\emph{However, the two models neglect many indispensable factors, such as delay time, topic, and network topology evolution.} 
For these deficiencies, there are a large number of extensions of those two models.
We briefly summarize the extensions in Table \ref{TAB:SummaryOfProgressive} and discuss them in detail in the following context.

\begin{table*}[!h]
  \scriptsize
  \caption{\label{TAB:SummaryOfProgressive}Summary of IC/LT-based Extension Models}
  \newcommand{\tabincell}[2]{\begin{tabular}{@{}#1@{}}#2\end{tabular}}
  \centering
  \begin{tabular}{ p{1.2cm} p{2cm} p{4.2cm} p{4.2cm} }
    \toprule
    \tabincell{c}{\textbf{Extensions}} &\tabincell{c}{\textbf{Articles}}    & \tabincell{c}{\textbf{Approaches}} &  \tabincell{c}{\textbf{Descriptions}} \\ \midrule
    \tabincell{c}{Delay}               &\tabincell{c}{\cite{saito2011learning, guille2012predictive}}    & \tabincell{c}{\(r_{uv}^{delay}\Longrightarrow\Delta \mathcal{I}(t+\lceil r_{uv}^{delay} \rceil)\)} & Add delay parameters \(r_{uv}^{delay}\) for each edge. Therefore, user activation is not restricted to consecutive steps \\ \midrule
    \tabincell{c}{Activation\\Pattern}    &\tabincell{c}{\cite{gruhl2004information, chen2012time, wang2017spatial, wang2017sadi}}    & \tabincell{c}{\(\mathbb{I}(\boldsymbol{p}_{u}(t))\Longrightarrow f(\boldsymbol{p}_{u}(t), \boldsymbol{r}_{u}(t),\cdots)\)} & Define more 'detailed' probabilities (\textit{i.e.}, reading probability, meeting probability) and modify the form of the activate function. \\ \midrule
    \tabincell{c}{User Class}          &\tabincell{c}{\cite{serrano2015novel, jiang2016rumor}}   & \tabincell{c}{\(\Delta \mathbb{I}(t)\Longrightarrow\sum_{\mathcal{S}(t)}\mathbb{I}(t)\pm\sum_{\mathcal{I}(t)}{\mathbb{I}^{'}(t)}\)} & Introduce more user classes. Then add the number of users who have switched from other classes to infected, and minus the number of users who have switched from infected to others. \\ \midrule
    \tabincell{c}{Semantic}            &\tabincell{c}{\cite{he2012influence,barbieri2013topic,lu2015competition,li2019modeling}}   & \tabincell{c}{\(p_{uv}\Longrightarrow p_{uv}^{c}\)}          &   Assign topic distribution to each user and edge. Only topic-matched information can be propagated through the social link. \\ \midrule
    \tabincell{c}{Dynamic\\Topology}    &\tabincell{c}{\cite{kempe2005influential, zarezade2017correlated, farajtabar2015coevolve,xie2015dynadiffuse}}   & \tabincell{c}{\(p_{uv}\Longrightarrow p_{uv}(t)\)}                 &   Make the network topology evolves with information diffusion. \\ \midrule
    \tabincell{c}{Multi-\\channels}      &\tabincell{c}{\cite{myers2012information, sahneh2013generalized, zhang2016information}}   & \tabincell{c}{\(f(\boldsymbol{p}_{u}(t))\Longrightarrow f(p_{u}^{1}(t))\diamond,\ldots,\diamond f(p_{u}^{s}(t))\) \\ \(\diamond \in \left\{\lor,\land,+, \ldots\right\}\)} &  Consider information spread across multiple platforms. \\ \bottomrule
  \end{tabular}
\end{table*}

\emph{Delay:}
Extensions for delay make propagations be no longer restricted to discrete consecutive time ticks. 
In addition, the user activated at time \(t\) is modified to \(\Delta \mathcal{I}(t+\lceil r_{uv}^{\text{delay}} \rceil)\).
AsIC (Aysnchronous Independent Cascades) and AsLT (Aysnchronous Linear Threshold) \cite{saito2011learning} add delay parameters \(r_{uv}^{delay}\) to each edge \((u,v)\). 
Compared with IC/LT, various delay parameters allow the user's neighbors to be activated at different time \(t\).
T-Basic (Time-Based Asynchronous Independent Cascades) \cite{guille2012predictive} expands fixed delay parameters into time-dependent functions, which can describe the dynamics of diffusion processes more accurately.

\emph{Activation Pattern:}
Extensions of activation patterns aim at devising a more elaborate activation function. 
Kempe \textit{et al.} \cite{kempe2005influential} proposed General Threshold (GT) Model which defines a general activation function \(f_{v}(N_{v}^{in}, S)\) for each user \(v\), instead of using an identical accumulative approach like the LT model. 
Chen \textit{et al.} \cite{chen2012time} introduced the ``meet probability'' \(r_{uv}^{meet}\) to indicate the possibility of contact between two users. 
The user will become active with the joint probability of ``meet probability'' \(r_{uv}^{meet}\) and the activation probability \(p_{uv}\).  
The activate function is defined as 

\begin{equation}
  p_{v}(t) = \begin{cases}
    1 & \text{If}\  rand_1>p_{uv} \land rand_2>r_{uv}^{meet},\ldots \\
    0 & \text{Otherwise}
  \end{cases}
  \label{EQU:MeetProbability}
\end{equation}
In addition, there are some other similar improvements (\textit{e.g.}, reading probabilities \cite{gruhl2004information}, message notification, and message-forgetting mechanisms \cite{wang2017spatial, wang2017sadi}) can be seen as the refinement of Eq. \ref{EQU:MeetProbability}. 
The target user will be activated when all these conditions are satisfied. 

\emph{User Class:}
User class extensions introduce new user classes based on IC/LT models compatible in some specified scenarios. 
Such as rumor spreading, since some susceptible users may become anti-rumors after receiving a rumor.  
Serrano \textit{et al.} \cite{serrano2015novel} proposed a rumor spreading model and they defined four types of users: neutral (initial state, inactive), infected 
(believe the rumor, active), vaccinated (believe the anti-rumor before being infected), and cured (believe the anti-rumor after being infected). 
It assumes that neutral users may become vaccinated with probability \(p_{uv}^{MakeDenier}\), and vaccinated users attempt to cure or vaccinate their neighbors with probability \(p_{uv}^{AcceptDeny}\).
Jiang \textit{et al.} \cite{jiang2016rumor} assumed that every susceptible user (S) can be infected (I) by their neighbors with probability \(p_{uv}\), and then be recovered (R) with probability \(q_{u}(t)\). 
To more precisely describe user states in rumor spreading, they introduced two sub-state of infected users: `contagious' (Con) and `misled' (Mis). 
An infected user first becomes contagious and then transits to be misled the next time \(t\). 
This mechanism can reflect the sequence of user infection during the rumor spreading, which is essential for source identification. 
Therefore, at time tick \(t\), the actual number of newly infected users should not include those who switched from infected into other states.

\emph{Semantic:}
Existing semantic extensions primarily consider the impact of topics on the diffusion process. 
For different topics, the main idea of these models \cite{he2012influence,barbieri2013topic,lu2015competition, li2017modeling,li2019modeling} is to \emph{assign topic distributions to each user and topic-related propagation probability to each edge}. 
For example, Zhu \textit{et al.} \cite{zhu2016minimum} assumed that every user \(v\) adopts each piece of the information with equal probability. 
He \textit{et al.} \cite{he2012influence} thought that every user has three states, \(inactive\), \(active^+\), and \(active^-\), with two independent threshold, positive \(\theta_{v}^{+}\) and negative threshold \(\theta_{v}^{-}\). 
Each edge \((u,v)\) has two weights, positive \(w_{uv}^{+}\) and negative weight \(w_{uv}^{-}\). 
For every inactive user \(v\), two types of influence are accumulated separately.
If \(\sum_{u\in\mathcal{N}_{I}(v)}{w_{uv}^{+}\geq\theta_{v}^{+}}\) is satisfied, the user \(v\) will switch to \(active^+\), otherwise will switch to \(active^-\). 
Com-IC \cite{lu2015competition} introduces the reconsideration mechanism in which one user may adopt the opinion \(B\) if he has adopted the complementary opinion \(A\). 
Li \textit{et al.} \cite{li2017modeling} utilized the payoff matrices to calculate user's payoffs, thereby deciding social choice for each piece of information.  
It should be noted that although these studies focus on the diffusion of multiple pieces of information, the majority of them stipulate that the user will not adopt the opposite information after they have been activated.

\emph{Dynamic Topology:}
It has been confirmed that information diffusion will promote network evolution \cite{li2016exploiting}. 
Dynamic topology extensions change propagation probability between user-pair from \(p_{uv}\) (fixed) to \(p_{uv}(t)\) (time-varying). 
Decreasing Cascades (DC) \cite{kempe2005influential} thinks that the probability between users \(u\) and \(v\) diminish as their active neighbor number increases. 
Zarezade \textit{et al.} \cite{zarezade2017correlated} used Hawkes process to model users' behavior adoption intensity (probability). The \(\lambda_{u}(t,act)\) defines the intensity of action \(act\) taken by the user \(u\) at time \(t\), and it can be expressed as \(\lambda_{u}(t,act)=\lambda_{u}(t)f_{u}(act|t)\). 
The \(\lambda_{u}(t)\) represents the intensity of the action taken by the user \(u\) at time \(t\), 
and \(f_{u}(act|t)\) represents the intensity that the user \(u\) takes the action \(act\) on the premise of time \(t\). 
DynaDiffuse \cite{xie2015dynadiffuse} models the dynamic characteristics in an ingenious way.
If a new connection appears, it assigns a small positive value to the newly created link. 
When the edge rate exceeds this positive value, the edge (re)emerges, and conversely, the edge disappears.
When all edges of a user disappear, the user can be considered as removed. 
In recent years, scholars introduced user interaction in network evolution. 
Co-Evolve \cite{farajtabar2015coevolve} considers two types of events: tweet/retweet events, \(e^{r}\), and link creation events, \(e^{l}\) as triplets:
\begin{equation*}
  e^{r} \text{ or   } e^{l} := (u, s, t)
\end{equation*}
The triplet indicates that the event when the source user \(s\) propagates a message to the target user \(u\) or establishes a friendship link with the target user \(u\) at time \(t\).
Next, researchers designed a multivariate Hawkes process to capture the intensity of user boosted by a previous diffusion process, and modeled the link creation using the survival process. 
After that, they coupled the intensity of link creation with retweeting events.

\emph{Multi-channels:}
Multi-channels extensions attempt to study information diffusion in multiple social platforms. 
Because the ``closed-world'' restriction is far from reality as users can receive information from various channels such as TV and telephone. 
Myers \textit{et al.} \cite{myers2012information} thought that the arrival time of external exposures for each user follows a binomial distribution. 
Finally, the user activation probability is multiplied by the probability of the user is activated by internal and external influences.
Muse (Multi-source Multi-channel Multi-topic diffUsion SElection) model \cite{zhang2016information} studies the information diffusion across online Enterprise Social Networks (ESN) and offline organizational structure. 
Information propagates through online, offline, and hybrid (online and offline) diffusion channels among employees. 
Different diffusion channels will be weighted based on their importance learned from the social activity log data with optimization techniques. 
The user \(v\) will become activated if the intensity he receives from these three channels meets the activation threshold. 
Sahneh \textit{et al.} \cite{sahneh2013generalized} embeds the diffusion process in multi-layer networks where all networks have the same users but different edges. 
Moreover, they applied the GEMF (Generalized Epidemic Mean-Field) approximation to reduce the state space.
Generally, Muse \cite{zhang2016information} can be thought of as a particular LT model applied in multi-channels, because a user will become activated once the total threshold is exceeded. 
In comparison, Myers \textit{et al.} \cite{myers2012information}, and GEMF \cite{sahneh2013generalized} are the IC model applied in multi-channels, as a user will be activated as long as he is activated at any layer.   

\textbf{Pros and Cons:} Progressive models are intuitive to expand. 
They only need to know the network topology to predict the information diffusion process. 
Furthermore, their monotonicity and submodularity \cite{TR} make them widely used in some downstream application tasks, such as influence maximization and source identification. 
However, limitations of progressive models are apparent. 
First, their computational efficiency is very low. 
Despite progressive models omit many details of diffusion, such as the interaction of multiple pieces of information, the IM problem based on progressive models is \textit{NP-hard} \cite{TR}. 
Second, their performance is highly dependent on the completeness and accuracy of network topology, which is almost impossible to be satisfied in the real world. 
Finally, the predictive performance of these models depends on the description of user interaction behavior. 
For example, Correlated Cascades \cite{zarezade2017correlated} achieved significant performance improvements on multiple datasets by considering the interaction between users and the propagation process. 
Myers \textit{et al.} \cite{myers2012information} found that found that external links activated at least 29\% of users by analyzing the URLs forwarded by users.  
As more social behavior characteristics are discovered, it means that such models have enormous space for improvement.

\subsubsection{Non-progressive Models}
\label{sec:TDM:IAM:NonProgressive}
Progressive models assume that the user remains active after being activated without regret. 
However, in the real world, people may change their minds after communicating with friends. 
Therefore, non-progressive models introduce the \textbf{re-adoption} mechanism, and they are similar to SIS models.

\textbf{Assumptions:}
Compared with above progressive models, in non-progressive models, a user can be activated again by another piece of information \(I_{message}^{'}\) after he was activated by the \(I_{message}\). 

\begin{algorithm}[h]
  \caption{Basic Ideas of Non-Progressive Models}
  \label{ALG:WorkflowOfNonProgressive}
  \LinesNumbered
  \KwIn{$G=(V,E)$: Social topology; $\mathcal{I^i}(t)$: User set with \emph{i}-th opinion at time $t$;$T$: Deadline; }
  \KwOut{$\mathcal{I}(T)$: Opinion distribution at time $T$}

  \For{$u$ in $V$}{
    Initialize $u$'s opinion\;
  }

  $t = 0$\;
  \While{$t<T$}{
    \For{$u$ in $V$}{
      Choose one of $u$'s neighbors to update his opinions based on payoffs \;
    }
  }

\end{algorithm}

\textbf{Methods:}
Algorithm \ref{ALG:WorkflowOfNonProgressive} is the basic idea of non-progressive models. 
Supposing there are \(K\) topics, the probability that user \(u\) adopts topic \(c\) at time \(t\) is determined by the user's payoff: 

\begin{equation*}
    p_{u}^{c}(t)=\frac{payoff_{u}^{c}(t)}{\sum_{i\in K}payoff_{u}^{i}(t)}
\end{equation*}
The payoff refers to the attention, retweeting, reputation, and other things that the user expects to get after re-sharing/rejecting the message. 

Voter model \cite{clifford1973model} is an early typical non-progressive model in which the proportion of neighbors' opinions determines the user's payoff.  
In other words, the user \(v\) randomly selects one of his neighbor \(u\) to communicate and turns his point of view into that of the user \(u\)'s. 
Li \textit{et al.} \cite{doi:10.1080/15427951.2013.862884} divided user neighbors into positive and negative ones (\textit{e.g.}, enemy, foe). 
Therefore, at each step, users are more inclined to adopt the opinions of the majority of users. 
The authors proved that under different network structures, the steady-state (opinion) of distribution is different. 
Kimura \textit{et al.} \cite{10.1007/978-3-642-33486-3_36} introduced temporal decay into voter models. 
Specifically, when computing the probability that the user $u$ chooses the $i$-th opinion at time $t$, it is necessary to consider the influence of all opinions of the neighbor adopting within a certain time window. And the influence can be modeled via decay functions (\textit{e.g.}, power-law, exponential). 
In fact, game theory can also be used to calculate user payoffs \cite{jiang2014graphical, jiang2014evolutionary}. However, due to its high complexity, researchers usually assume that a group of users use the same revenue matrix.

Friedkin and Johnsen (F-J) \cite{Friedkin_Johnsen} considered user stubbornness (or self-persistence) into opinion dynamics. 
Initially, each user $u$ is assigned with a opinion $x_u(0)$. 
During each round of social interaction, the user may update the opinion from neighbors with probability $\theta_u$, or stick to the original opinion with probability $1-\theta_u$. 
Ghaderi \textit{et al.} \cite{GHADERI20143209} expressed the user's opinion as a value between 0 and 1. 
And they derived that the convergence bound depends on the network structure, the location of stubborn users, and their stubbornness. 
Tian \textit{et al.} \cite{TIAN2018213} believed that path-dependent (\textit{i.e.,} repeatedly arising or interdependent) topic might achieve consensus, and the opinion evolution will enhance the network connectivity. 

Most of the above models assume that an individual user will only hold a single opinion simultaneously.  
However, in the real world, vast opinions or topics are not black and white, such as the centrist phenomenon \cite{RUSINOWSKA20199}. 
Researchers considered that users could hold multiple opinions simultaneously to improve model generality. 
Generally, these opinions have different strengths and satisfy a particular distribution. For example, Nayak \textit{et al.} \cite{nayak2019smart} utilized the Dirichlet distribution to describe opinion distributions of users, and the Bayesian method was employed to update the opinion distribution after receiving new messages.

\begin{table}
  \footnotesize
  \caption{\label{TAB:SummaryOfNoProg}Summary of Non-Progressive Models}
  \newcommand{\tabincell}[2]{\begin{tabular}{@{}#1@{}}#2\end{tabular}}
  \begin{tabular}{ p{1.5cm} p{3.2cm}  p{6.8cm} p{1cm}}
    \toprule
    \tabincell{c}{\textbf{Extensions}} &\tabincell{c}{\textbf{Update}}    & \tabincell{c}{\textbf{Explanations}} & \tabincell{c}{\textbf{Articles}}\\ \midrule

    \tabincell{c}{Voter \\ (Original)} & \tabincell{c}{$p_u^{i \leftarrow j}(t) = \frac{d_u^j(t)}{d_u(t)}$} & \tabincell{l}{User update their opinions based on the proportion \\of neighbors' opinions.}   & \tabincell{l}{\cite{clifford1973model}}   \\ \midrule

    \tabincell{c}{Voter \\ (Signed\\Network)} & \tabincell{c}{$x_u(t) = \sum_{j \in V}\frac{A^{+}_{uj}}{d_u}x_j(t-1) $\\$+ \sum_{j \in V}\frac{A^{-}_{uj}}{d_u}(1-x_j(t-1))$} & \tabincell{l}{Divide users into positive and negative groups. \\When the user $i$ receives more positive influence, \\he will become positive, and vice versa. $x_u(t)$ is the\\ probability that the user $u$ is positive at time $t$. }   & \tabincell{l}{\cite{doi:10.1080/15427951.2013.862884}}   \\ \midrule

    \tabincell{c}{Temporal\\Decay} & \tabincell{c}{$p_u^k(t) = \frac{1+\sum_{j\in D(u)}|M_k(t,j)|}{K+\sum_{j\in D(u)}|M_k(t,j)|}$} & \tabincell{l}{Divide users into $K$ groups according to their opinions. \\$M_k(t,j)$ is the set of past opinion $k$'s influence that the \\user $j$ receive at time $t$. $p_u^k(t)$ is the probability that the \\user $u$ adopts the opinion $k$ at time $t$.}   & \tabincell{l}{\cite{10.1007/978-3-642-33486-3_36}}   \\ \midrule

    \tabincell{c}{F-J} & \tabincell{c}{$x_u(t) = \theta_u x_u(t) $\\$+(1-\theta_u) x_u(0)  $} & \tabincell{l}{During each interaction, the user $u$ is stubborn to his \\initial opinion with probability $1-\theta_u$.  }   & \tabincell{l}{\cite{GHADERI20143209, TIAN2018213}}   \\ \midrule

    \tabincell{c}{Bayesian}  &\tabincell{c}{$p_u^k(t) = \beta_u p_u^k(t) + \gamma_u n_u^k(t)$} & \tabincell{l}{Users can have multiple opinions, and these opinions \\follows the Dirichlet distribution. User's opinion can \\be updated based on Bayesian method. }& \tabincell{l}{\cite{nayak2019smart}}  \\ \bottomrule

  \end{tabular}
\end{table}

\textbf{Pros and Cons:} 
Compared with progressive models, non-progressive ones introduce the ``re-adoption'' mechanism, making them more suitable for modeling multiple pieces of information diffusion scenarios. 
The authenticity is enhanced. 
However, these models do usually take more time to reach convergence than those progressive ones. 
Further, due to the introduction of the ``re-adoption'' mechanism, non-progressive models do not have monotonicity and submodularity, which means that the infected user set's influence may diminish as its size increases. 
Roughly, voter models reflect the fact that the majority population influences the user's opinion, and to some extent, demonstrates the process of user opinion evolution \cite{sichani2017inference}. 
In the short term, the signed social vectors will increase the spreading speed by 20\% to 40\% \cite{doi:10.1080/15427951.2013.862884}. 
Unlike influence maximization, if the initial infected users are randomly selected, the volume of infection will become smaller as the opinions evolve \cite{nayak2019smart}. 
Non-progressive models are essential to the research of opinion evolution, and have broad prospects.

  \begin{table}
    \newcommand{\tabincell}[2]{\begin{tabular}{@{}#1@{}}#2\end{tabular}}
    \caption{\label{TAB:SummaryOfDeepIndividual}Summary of Deep Learning (Individual) Models}
    \scalebox{0.8}{
      \renewcommand\arraystretch{1}
      \begin{tabular}{p{1.8cm} p{3cm} p{6.5cm} p{2.8cm}}
        \toprule
        \tabincell{c}{\textbf{Scopes}} &  \tabincell{c}{\textbf{Key Components}} & \tabincell{c}{\textbf{Contributions}}  & \tabincell{c}{\textbf{Articles}} \\ \midrule
  
        \tabincell{c}{Content \& \\User-related} &  \tabincell{c}{UA-CNN \\(CNN + Attention \\+ Similar Matrix)} & \tabincell{l}{CNN: Embed tweet contents \\ Attention: Embed user interest \\ Similar Matrix: Compute the probability of \\user retweet}  & CIKM 2016 \cite{zhang2016retweet}\\ \midrule

        \multirow{7}{*}{\tabincell{c}{Structural \\ \& Temporal }} &  \tabincell{c}{Topo-LSTM\\(Embedding \\+ Topo-LSTM \\+ Mean Pooling)} & \tabincell{l}{Embedding: Embed static preference and \\dynamic context of sender in cascade graph\\ Topo-LSTM: Learn activation relationships \\among senders and receivers. \\ Mean Pooling: Fuse multiple senders}  & ICDM 2017 \cite{wang2017topo}\\ \cmidrule{2-4}

        &  \tabincell{c}{DeepDiffuse\\(CAN + CPN)} & \tabincell{l}{CAN: Embed cascades through the \\combination of LSTM, Attention and other \\components \\ CPN: Predict the next active user and its \\activation time through LSTM, FFN,\\and Softmax }  & ICDM 2018 \cite {DBLP:conf/icdm/IslamMAPR18}\\ \cmidrule{2-4}

         &  \tabincell{c}{DeepInf\\(Embedding + \\GCN + GAT)} & \tabincell{l}{Embedding: Embed users (include social \\influence locality \cite{zhang2013social}) \\ GCN/GAT: Embed social network structure }  & KDD 2018 \cite{qiu2018deepinf}\\ \cmidrule{2-4}

         &  \tabincell{c}{DMT-LIC\\(GAT + Bi-LSTM \\+ Shared-Gate)} & \tabincell{l}{GAT + Bi-LSTM: Embed cascades graph from \\cascade-level (structural) and user-level\\ (sequential), respectively \\ Shared-Gate: Fuse user importance, sequential, \\and structural features }  & SIGIR 2019 \cite{DBLP:conf/sigir/ChenZ0TZZ19}\\ \cmidrule{2-4}

         &  \tabincell{c}{FOREST\\(GCN + GRU + RL)} & \tabincell{l}{GCN: Embed structural context \\ GRU: Model sequential cascades data \\ RL: Guide user-level prediction by predicted \\cascade size }  & IJCAI 2019 \cite{yang2019multi}\\ \cmidrule{2-4}

        &  \tabincell{c}{CoupledGNN\\(State GNN + \\Influence GNN)} & \tabincell{l}{State GNN: Model the user activation state \\ Influence GNN: Model the interaction \\between users }  & WSDM 2020 \cite{cao2020popularity}\\ \cmidrule{2-4}

         &  \tabincell{c}{NDM\\(Embedding + \\Attention + CNN)} & \tabincell{l}{Embedding: Embed users \\ Attention: Learn the activation relationships \\among active users \\ CNN: Capture the decay effect of users \\activated at different positions }  & TKDE 2021 \cite{DBLP:journals/tkde/YangSLHLL21}\\ \midrule

        \tabincell{c}{Temporal \\ \& Content  \& \\ User-related } &  \tabincell{c}{MMVED\\(MLP + LSTM)} & \tabincell{l}{MLP: Encode multi-modal information for \\variational inference \\ LSTM: Encode and decode  sequential cascades }  & WWW 2020 \cite{DBLP:conf/www/XieZZPYHLC20}\\ \midrule

        \tabincell{c}{Structural \\ \& Content  \& \\ User-related}  &  \tabincell{c}{DiffNet\\(Embedding + MLP\\ + Influence Diffusion \\Layer)} & \tabincell{l}{Embedding: Embed users and items \\ MLP: Fuse associated features \\ Influence Diffusion Layer: Embed user \\influence by diffusion process }  & SIGIR 2019 \cite{wu2019neural}\\ \cmidrule{2-4}

      &  \tabincell{c}{Inf-VAE\\(VAE + \\Positional Encoding + \\Co-attentive Fusion)} & \tabincell{l}{VAE: Model social homophily \\ Positional Encoding: Model temporal influence \\ Co-attentive Fusion: Fuse the above two \\features to generate future cascade }  & WSDM 2020 \cite{sankar2020infvae}\\ \bottomrule

      \end{tabular}
      }
  \end{table}

\subsubsection{ML-based (Individual) Models}
\label{sec:TDM:IAM:FeatureBased}
Compared with ML-based models introduced in Section \ref{sec:TDM:DVM:Feature}, models in this section aim at predicting which user will be infected. 

\textbf{Assumption:} Similar to models in Section \ref{sec:TDM:DVM:Feature}, ML-based (individual) ones relax almost all unrealistic assumptions. 
Nevertheless, their performance depends on whether the algorithm can effectively learn the propagation characteristics.

\textbf{Methods:}
Generally, modeling techniques can be roughly divided into two groups: \textbf{feature-based} and \textbf{deep learning}. 
For \textbf{feature-based} models, besides the native machine learning algorithms introduced in Section \ref{sec:TDM:DVM:Feature}, the collaborative filtering techniques in the recommendation system can also be used to predict individual adoption, such as matrix factorization \cite{ma2008sorec, ma2011recommender, jiang2012social, jiang2015message, cui2011should}, tensor decomposition \cite{jiang2014fema, hoang2017gpop, hoang2016microblogging}, and transfer learning \cite{jiang2016little}. 
In essence, these models try to find out the retweet matrix \(R\) based on user matrix \(U\) and message matrix \(V\) where 
$$R \approx U^{T}V$$
Tensor \(R\), \(U\), and \(V\) are learned from historical interaction records.

Similar to deep learning (volume) models in Section \ref{sec:TDM:DVM:Feature}, the framework of \textbf{deep learning} (individual) ones consists of four major components: embedding data as vectors, using neural network models to learn diffusion features, aggregating features, and prediction, 
Generally, the differences between these two types of models are summarized as follows. 
\textit{First}, volume models use MLP to predict the diffusion volume $I(t)$, while individual ones mostly use the Softmax classifier to get the activation probability of each user $\alpha_{uv}(t)$. 
\textit{Second}, these models must be able to learn individual-level features, while volume ones do not need to meet this requirement. 
Even if user-related features are required, only the content publishers are needed \cite{DBLP:conf/www/ZhangWWZ18, DBLP:conf/dasfaa/WangZW18}. 

For content and user-related features, Zhang \textit{et al.} \cite{zhang2016retweet} integrated attention mechanism and CNN to learn relationships among user interests and content, and utilized the similarity matrix to infer future forwarding users. 
Xie \textit{et al.} \cite{DBLP:conf/www/XieZZPYHLC20} employed MLP as VAE (Variational Auto Encoder) \cite{DBLP:conf/nips/wu18multimodal} to generate the hidden representations of multi-modal information, and taken cascade sequence and social information hidden representations as input into LSTM \cite{DBLP:journals/neco/HochreiterS97} to predict the future active users. 
Wu \textit{et al.} \cite{wu2019neural} devised a layer-wise model to track user latent embeddings evolving as the social diffusion process continues, and predicted user adoption probability with the fusion of users and items features. 
Sankar \textit{et al.} \cite{sankar2020infvae} modeled users' social homophily and temporal features by VAE and Positional Encoding \cite{DBLP:conf/nips/vaswani17attentionis}, respectively. 
And then, they developed a co-attentive fusion mechanism to fuse the above two types of features. 

For structural and temporal features, roadmaps can be roughly classified as follows: \textbf{NodeEmbedding+RNN}, \textbf{GNN+RNN}, \textbf{NodeEmbedding+GNN}, and others.  
For \textbf{NodeEmbedding+RNN} models, they aim to make forwarding prediction directly through the encoding and decoding of the cascade graph through the Encoder-Decoder architecture. 
Wang \textit{et al.} \cite{wang2017topo} proposed Topo-LSTM, which takes node embedding and cascade graphs as input to learn cascade structural and temporal features. 
\textbf{GNN+RNN} models extract the structural features of cascade graph or network topology through the GNN model, and then learn the sequential features through the RNN model. 
Chen \textit{et al.} \cite{DBLP:conf/sigir/ChenZ0TZZ19} embedded cascade graph (structural) and user-level sequential (temporal) features via GAT (Graph Attention Networks) \cite{DBLP:conf/iclr/VelickovicCCRLB18} and Bi-directional LSTM, respectively. 
The shared-gate mechanism is developed to fuse user importance, cascade-level, and user-level features.   
Yang \textit{et al.} \cite{yang2019multi} introduced reinforcement learning, guiding individual prediction through the macro volume prediction.  
For  \textbf{NodeEmbedding+GNN} models, they predict future forwarding through user relevance. 
Qiu \textit{et al.} \cite{qiu2018deepinf} modeled the strengths of neighbor relationships via social influence locality \cite{zhang2013social}. 
After that, GCN \cite{DBLP:conf/iclr/KipfW17} and GAT \cite{DBLP:conf/iclr/VelickovicCCRLB18} are employed to learn network structural features for user relevance extraction. 
In this article, temporal features are implicitly encoded in the social influence locality.
Cao \textit{et al.} \cite{cao2020popularity} devised StateGNN and InfluenceGNN to model user states and activation influence, respectively. 
In this way, through the graph neural network propagation and aggregation mechanism, the interplay of the two graphs is realized to "simulate" the information propagation, and temporal features can be captured through iterative interaction. 
Besides, Yang \textit{et al.} \cite{DBLP:journals/tkde/YangSLHLL21} utilized the attention mechanism \cite{DBLP:conf/nips/vaswani17attentionis} to learn user activation relationships, coupling the CNN to capture the decay effect of active users.

\textbf{Pros and Cons:}
Similar to the advantages and disadvantages of volume ML-based models, individual ML-based ones do not need to design complex mathematical expressions, while they require computational expensive feature engineering or learning representation to extract features. 
Compared with the volume models, the individual model needs more fine-grained features. 
Taking user-related features as an example, volume models only need publisher features \cite{DBLP:conf/dasfaa/WangZW18, DBLP:conf/www/ZhangWWZ18}, while individual models require features of all potential retweet users.  
Besides, their model performance is highly dependent on the quality and quantity of features. 
The model's prediction performance with whole-structure (friendship networks and cascade structure) input \cite{qiu2018deepinf, yang2019multi} is an order of magnitude higher than that of the model with partial structure (cascade structure) input \cite{sankar2020infvae, DBLP:journals/tkde/YangSLHLL21}.

\begin{table}[!htbp]
  \caption{\label{TAB:SummaryOfIndividualModels}Summary of Individual Adoption Models}
  \newcommand{\tabincell}[2]{\begin{tabular}{@{}#1@{}}#2\end{tabular}}
  \centering
  \scalebox{0.85}{
  \begin{tabular}{p{1.8cm} p{2.2cm} p{2cm} p{2.4cm} p{2cm} p{2.5cm}}
    \toprule
    \multicolumn{2}{c}{}& \tabincell{c}{\textbf{Assumptions}} & \tabincell{c}{\textbf{Monotonicity} \&\\\textbf{Submodularity}}  &\tabincell{c}{\textbf{Explicit}\\\textbf{Expressions}}  &\tabincell{c}{\textbf{Key Factors}} \\
    \midrule 
    \multirow{2}{*}{\tabincell{c}{Time-series}} & \tabincell{c}{Progressive} & \tabincell{c}{$S\rightarrow I$} & \tabincell{c}{$\surd$}   & \tabincell{c}{$\surd$} & \multirow{2}{*}{\tabincell{c}{Model rationality \\ \& Network topology}} \\ \cmidrule{2-5}

                                             & \tabincell{c}{Non-progressive} & \tabincell{c}{$S\rightarrow I \rightarrow S$} & \tabincell{c}{-}    & \tabincell{c}{$\surd$} &  \\ \midrule

    \multirow{2}{*}{\tabincell{c}{ML-based}}  & \tabincell{c}{Classification} & \tabincell{c}{-} & \tabincell{c}{-}    & \tabincell{c}{-} & \tabincell{c}{Features extracted \\ from datasets} \\ \bottomrule
  \end{tabular}
  }
\end{table}

\subsubsection{Analysis and Discussion}
The summary of individual adoption models is shown in Table \ref{TAB:SummaryOfIndividualModels}. 
There are four main characteristics of progressive and non-progressive models. 
\textit{First}, both rely on social topology as they all assume that information is spreading on social networks via social links. 
\textit{Second}, progressive models assume infected users remain stable once they were infected without any regrets. 
On the contrary, any user may shift his grounds multiple times based on his payoffs in non-progressive models.  
\textit{Third}, in each activation iteration of progressive models, only those inactive can be selected as a newly active user, whereas every user should be considered under non-progressive models. 
\textit{Finally}, non-progressive models' monotonicity and submodularity are broken because of their re-adoption mechanism. 
Although the time complexity of non-progressive models is higher, they are more in line with real-world diffusion scenarios, capturing opinion evolution dynamics. 

In terms of ML-based models, their advantages and disadvantages are similar to those illustrated in Section \ref{sec:TDM:DiffusionVolumeModels}. 

\subsection{Relationship Inference Models}
\label{sec:TDM:RelationshipInference}
Generally, researchers can infer the user-to-user propagation relationship by likelihood maximization (Time-series) and diffusion embedding approaches (Data-driven). 
Representative works are introduced as follows.

\subsubsection{Likelihood Maximization Models} 
\label{sec:TDM:RIM:ProbabilityInferene}

These models devise likelihood functions to infer propagation relationships among users based on their interaction history of cascades. 

\begin{table}[H]
  \scriptsize
  \caption{\label{TAB:SummaryOfLikelihood}Summary of Likelihood Maximization}
  \newcommand{\tabincell}[2]{\begin{tabular}{@{}#1@{}}#2\end{tabular}}
  \begin{tabular}{ p{1.5cm} p{1.5cm}  p{3.5cm} p{5cm}}
    \toprule
    \tabincell{c}{\textbf{Types}} &\tabincell{c}{\textbf{Articles}}    & \tabincell{c}{\textbf{Approaches}} & \tabincell{c}{\textbf{Explanations}}\\ \midrule

    \tabincell{c}{Discrete \\ Time} & \tabincell{c}{\cite{gomez2012inferring,gomez2012submodular,xu2018contrastive, DBLP:conf/icde/HanTZHHG20}} & \tabincell{c}{$\begin{aligned}  L(\textbf{A}; \mathcal{C})&=\prod_{c \in \mathcal{C}} P(Tree;c)\\  &=\prod_{c \in \mathcal{C}} \prod_{(i,j)\in Tree}P_c(i,j)\end{aligned}$} & \tabincell{l}{$P(Tree;c)$: The probability that the diffusion \\trajectory of the cascade $c$ is a tree pattern $Tree$. \\ $p_c(i,j)$: The propagation probability between \\the user $i$ and $j$.}    \\ \midrule

    \tabincell{c}{Continuous \\ Time} & \tabincell{c}{\cite{CONNIE_myers2010convexity,rodriguez2011uncovering, gomez2013structure}\\\cite{ wang2012feature, wang2014mmrate}} & \tabincell{c}{$\begin{aligned}L(\textbf{A}; \mathcal{C}) = &\prod_{c \in \mathcal{C}} \prod_{t_i^c \leq T^c} \Gamma_i^{+}(\textbf{t}^c) \\ & \times \prod_{c \in \mathcal{C}}    \prod_{t_i^c \leq T^c} \Gamma_i^{-}(\textbf{t}^c)    \end{aligned}$} & \tabincell{l}{$\textbf{t}^c$: The vector recording the time of cascade $c$. \\ $ T^c$: Last time tick of the observable cascade $c$. \\ $\Gamma_i^{+/-}$: The probability that all active (inactive) \\users was activated (not activated) before time $T^c$}    \\ \midrule

  \end{tabular}
\end{table}

\textbf{Assumptions:}
To simplify the inference problem, researchers make two basic assumptions: 1) a user gets infected only once, 2) all the infections are conditionally independent.

\textbf{Methods:} 
As shown in Table \ref{TAB:SummaryOfLikelihood}, these models craft likelihood functions based on the observable cascades via discrete-time and continuous-time approaches, and maximize the likelihood function to infer the underlying diffusion relationships.

NetInf \cite{gomez2012inferring} and MultiTree \cite{gomez2012submodular} adopt \textbf{discrete-time} approaches to infer the propagation relationship through \textit{submodular optimization}. 
They both build a retweet tree for every cascade, and then select \(k\)-edges network based on these trees. 
The probability of observing cascade \(c\) propagating in a particular tree structure \(Tree\) can be written as:
\begin{scriptsize}
  \begin{equation}
    P(c|Tree)=\prod_{(i,j)\in Tree}p_c(i,j)
    \label{EQU:RetweetTreeMulti}
  \end{equation}
\end{scriptsize}
The propagation probability \(p_c(u,v)\) can be computed by three well-known models: Exponential, Power-law, and Rayleigh, with the time difference between user infection time \(\Delta=t_u^c-t_v^c\) as parameters. 
After building retweet trees for each cascade, \textbf{NetInf will find one most likely retweet tree and select \(k\) optimal edges from the tree as the diffusion network.} 
However, if the number of observable cascades is small, we may not find the appropriate retweet tree. 
Therefore, \textbf{MultiTree selects \(k\) optimal edges from all possible retweet trees}. 
Xu \textit{et al.} \cite{xu2018contrastive} regards network structure as a simple log-linear, edge-factored directed spanning tree. 
After that, an unsupervised, contrastive training procedure is utilized to infer the network structure. 
Han \textit{et al.} \cite{DBLP:conf/icde/HanTZHHG20} devised a mutual-information-based score criterion to find the propagation relationships, inferring the most probable diffusion network. 

CONNIE (Convex Network Inference) \cite{CONNIE_myers2010convexity} and NetRate \cite{rodriguez2011uncovering} adopt \textbf{continuous-time} approaches, and convert the network inference problem to \textit{convex optimization problem}.  
They all build a two-part likelihood function based on cascade records with the help of Survival theory. 
To expose the complicated concepts, we summarize the skeleton of continuous-time relationship inference models in Algorithm \ref{ALG:WorkflowOfContinuesTimeModels}. 
First, for every infected user \(v\) who was infected at time \(t_v^c\), they compute the probability that at least one previously infected user has infected him. 
Second, they compute the probability that none of the infected users activated him for every uninfected user. 
After that, network structure matrix \(\textbf{A}\) can be inferred by maximizing the likelihood function. 
NetRate utilizes survival theory to build the likelihood function and only considers infected users. 
As a result, NetRate models diffusion as a spatially discrete network of continuous, conditionally independent temporal processes. 

Compared with NetInf and MultiTree, CONNIE and NetRate can infer the \textbf{link strength} in addition to the \textbf{existence of link}.
However, the above four models can merely infer static network structures, and only the time factors are considered. 
Corresponding to the above problems, there are some improvement solutions based on NetRate, including InfoPath \cite{gomez2013structure}, MoNet \cite{wang2012feature}, MMRate \cite{wang2014mmrate}. 
InfoPath makes transmission rate \(p_{u,v}\) varying over time, and introduces cascades weights \(w_{c}(t)\)  based on the time interval to adjust by the cascade influences. 
These improvements make InfoPath support \textbf{dynamic network inference} problems. 
MoNet integrates the \textbf{additional features} \(\textbf{f}_{u}^{(i)}\) associated with each user into NetRate. 
For feature \(\textbf{f}_{u}^{(i)}\), if users \(u_1\), \(u_2\), \(u_3\) join the cascade at time \(t_1\), \(t_2\) and \(t_3\), and \(t_1 < t_2 < t_3\), then the Eq. \ref{EQU:MoNetFeature} is satisfied. 
\begin{scriptsize}
  \begin{equation}
    ||t_{u_3} - t_{u_2}||+||\textbf{f}_{u_3}-\textbf{f}_{u_2}||>||t_{u_3}-t_{u_1}||+||\textbf{f}_{u_3}-\textbf{f}_{u_1}||
    \label{EQU:MoNetFeature}
  \end{equation}
\end{scriptsize}
MMRate \cite{wang2014mmrate} focuses on the information diffusion in \textbf{multi-aspects(\textit{e.g.}, topics)}. 
User links have different strengths under different topics due to the variety of users' interests \cite{tang2012mtrust}. 
Hence, the transmission rate \(p_{u,v}\) and network topology \(G^{l}\) should be different under different topics. 

\begin{algorithm}[h]
  \caption{Basic Ideas of Continuous Time Models}
  \label{ALG:WorkflowOfContinuesTimeModels}
  \LinesNumbered
  \KwIn{$\mathcal{C} = \{c^1, \ldots,c^M\}$: Cascade set; \(c^n = \left\{ (u^n_1,t^n_1); (u^n_2, t^m_2);\ldots; \right\}\): The \emph{n}-th cascade ; $T^n$: Observable last time of the \emph{n}-th cascade}
  \KwOut{\(\textbf{A} = \{\alpha_{i,j}\}\): Matrix of transmission likelihood}

  $f(t^n_j|t^n_i, \alpha_{i,j}) \Leftarrow$ Devise conditional likelihood of transmission between the user $i$ and $j$ in the \emph{n}-th cascade\;
  $S(t^n_j|t^n_i; \alpha_{i,j})  = 1 - F(t^n_j|t^n_i;\alpha_{i,j})$ // Survival function: the probability that user $j$ is not activated by user $j$ by time $t_j$ in the \emph{n}-th cascade\;
  $H(t^n_j|t^n_i;\alpha_{i,j}) = \frac{f(t^n_j|t^n_i, \alpha_{i,j})}{S(t^n_j|t^n_i; \alpha_{i,j})}$ // Harzard function: instantaneous activate rate from the user $i$ to user $j$ in the \emph{n}-th cascade \;

  $f(\textbf{t}^n;\textbf{A}) = \prod_{t^n_j \le T^n} \prod_{t^n_m > T^n} S(T^n|t^n_j;\alpha_{j,m}) \times \prod_{k:t^n_k<t^n_j}S(t^n_j|t^n_k;\alpha_{k,j})\sum_{i:t^n_i<t^n_j}H(t^n_j|t^n_i;\alpha_{i,j})$ // Likelihood of the \emph{n}-th cascade\;

  $\text{minimize}_{\textbf{A}} -\sum_{c\in \mathcal{C}} f(\textbf{t}^c;\textbf{A})$ // Likelihood of cascade set

  \For{$c$ in $\mathcal{C}$}{
    \ForAll{(i,j):$t^c_i<t^c_j$}{
      Update $\alpha_{i,j}$
    }
  }

\end{algorithm}

\textbf{Pros and Cons:}
MultiTree \cite{gomez2012submodular} and NetInf \cite{gomez2012inferring} only care about the existence of the propagation relationship, and the edge number \(k\) is determined by experience. 
Compared with them, CONNIE \cite{CONNIE_myers2010convexity} and NetRate \cite{rodriguez2011uncovering} infer the existence and strength of link. Specifically, they do not need to set the number of edges. 
If \(p_{uv} \rightarrow 0\) is satisfied, it implies that there is no propagation relationship between user \(u\) and \(v\). 
But in terms of running time, MultiTree and NetInf are almost orders of magnitude faster than NetRate and CONNIE \cite{gomez2012submodular}. 
Subsequently, the stochastic-gradient optimization speeds up almost 10-100 times of continuous-time approaches \cite{gomez2013structure}. 
Besides, the introduction of individual characteristics will double the prediction performance \cite{wang2012feature}. 
Significantly, the greater the propagation probability between users, the higher the inference accuracy \cite{DBLP:conf/icde/HanTZHHG20}. 
However, these models rely on assumptions of independent and sole activation. Whether these assumptions are reasonable still needs to be verified.

\subsubsection{Diffusion Embedding Models}
\label{sec:TDM:RIM:Embedding}
The design philosophy of diffusion embedding models is to map the users and the observable information diffusion records into a unified and continuous semantic space. 
After that, space geometry knowledge is applicable to infer propagation relationships. 

\textbf{Assumptions:} Similar to the above ML-based models, diffusion embedding ones relax almost all unrealistic assumptions. 
Furthermore, their performance heavily depends on the quality of data and feature representation.

\begin{table}
  \caption{\label{TAB:SummaryOfEmbedding}Summary of Diffusion Embedding Models}
  \newcommand{\tabincell}[2]{\begin{tabular}{@{}#1@{}}#2\end{tabular}}
  \begin{tabular}{ p{2cm} p{1.5cm}  p{2.8cm} p{6cm}}
    \toprule
    \tabincell{c}{\textbf{Types}} &\tabincell{c}{\textbf{Articles}}    & \tabincell{c}{\textbf{Approaches}} & \tabincell{c}{\textbf{Explanations}}\\ \midrule

    \tabincell{c}{Probabilistic \\ Inference} & \tabincell{c}{\cite{bourigault2016representation, DBLP:conf/ispa/0006ZZ019}} & \tabincell{c}{$P^c_{u,v}=f^c(z_{u},z_{v})$} & \tabincell{l}{$P^c_{u,v}$: The possibility that the user $v$ forward \\the message $c$ published by the user $u$, and \\the function $f^c(\cdot)$ can be learned from data. \\ $z_{u}$: The embedding of the user $u$.  }    \\ \midrule

    \tabincell{c}{Space \\ Geometry} & \tabincell{c}{\cite{ bourigault2014learning,gao2017novel, DBLP:conf/dasfaa/SuZWFZY19}} & \tabincell{c}{$z_u + w_c\approx z_v$} & \tabincell{l}{$z_u$:  The embedding of the user $u$. \\ $w_c$: The embedding of the message $c$. }    \\ \midrule

  \end{tabular}
\end{table}

\textbf{Methods:} 
Existing diffusion embedding models can be classified into either \textbf{probabilistic inference} and \textbf{vector operation} models. 
For \textbf{probabilistic inference} models, they embed users into a unified semantic space, and the propagation probability of every user pair \((u,v)\) is computed by non-linear function \(f(z_{u},z_{v})\). 
After that, the Independent Cascade model simulates the diffusion process, and propagation relationships can be inferred via likelihood maximization. 
Bourigault \textit{et al.} \cite{bourigault2016representation} projected users into a continuous latent (euclidean) space, and used the Sigmoid function to calculate propagation probabilities. 
Finally, EM (Expectation-Maximization) algorithm is applied to infer the propagation relationship. 
Zhuo \textit{et al.} \cite{DBLP:conf/ispa/0006ZZ019} embedded uses via GAN (Generative Adversarial Nets) \cite{DBLP:conf/nips/GoodfellowPMXWOCB14}, avoiding complicated 
sampling with the help of generative-discrimination architecture.

For \textbf{space geometry} models, they embed users as well as messages into unified semantic space, and infer propagation relationships based on geometrical vector operations. 
Bourigault \textit{et al.} \cite{bourigault2014learning} projected users into a geometric manifold space where the source user is at the center of space, and other users are arranged in space according to their propagation relationships. 
Denoting embedding vectors' location of sender, receiver, and information as $z_u$, $z_v$ and $w_c$, the propagation relationship can be inferred based on the heat diffusion kernel $$
f(z_u, w_c, z_v, t) = (4 \pi t)^{\frac{n}{2}}e^{- \frac{||z_u+w_c-z_v||^2}{4t}}
$$. 
where $n$ is dimension. 
Gao \textit{et al.} \cite{gao2017novel} introduced TransE \cite{bordes2013translating} approach, which allows the inference can be achieved directly through vector operations as follows, omitting the design of diffusion kernel.  
$$
f_c(z_u, z_v)=||\textbf{W}_{c,u}\textbf{z}_{u}+\textbf{w}_{c}-\textbf{W}_{c,v}\textbf{z}_{v}||_{l-norm}
$$
where the information-specific matrices \(\mathbf{W}_{c, u}\in\mathbb{R}^{k\times k}\) identify the correlation between user \(u\) and information content \(c\). 
Su \textit{et al.} \cite{DBLP:conf/dasfaa/SuZWFZY19} embedded contents and users into heterogeneous information networks, and proposed the meta-path-similarity-based HWalk method to infer the diffusion network. 
The propagation probability between two nodes can be computed by the softmax function:
$$
p(u,v) = \frac{f(z_u, z_v)}{\sum_{j \in D_u^r} f(z_u, z_j)}
$$
where the $f(z_u, z_v)$ is the path similarity measurement, and $D_u^r$ is node type. 

\textbf{Pros and Cons:}
Compared with probabilistic models, vector operation ones are easier to do end-to-end training. 
And they make better use of the semantic-related features by embedding users and information into a unified space. 
However, similar to the ML-based models introduced before, the performance of diffusion embedding models relies on the data quantity and quality. 
Notably, the model performance decreases with the increment of content variance. 
According to the synthetic experiments, as the words of content increased from 5 to 50, model performance decreased by nearly 60\%. 
Moreover, this phenomenon cannot be solved simply by increasing the number of dimensions. 
Many researchers \cite{DBLP:conf/ispa/0006ZZ019, bourigault2014learning} have confirmed that blindly increasing the dimensionality may cause performance degradation due to overfitting.

\begin{table*}[!htbp]
  \tiny
  \caption{\label{TAB:SummaryOfNetworkInference}Summary of Propagation Relationship Inference}
  \newcommand{\tabincell}[2]{\begin{tabular}{@{}#1@{}}#2\end{tabular}}
  \centering
  \begin{tabular}{ p{2cm}||p{1cm} p{1cm} p{1cm}||p{1.2cm} p{1.2cm} p{1.2cm} p{1.2cm} }
    \toprule
    \multirow{2}{2cm}{Models} & \multicolumn{3}{c||}{Input} & \multicolumn{4}{c}{Support Inference} \\ 
    \cmidrule{2-8}
                                               & \tabincell{c}{Time}        & \tabincell{c}{User \\ Attribute} & \tabincell{c}{Information \\ Content} & \tabincell{c}{Transmission \\ Relationship} & \tabincell{c}{Transmission\\ Rate} & \tabincell{c}{Dynamic\\ Transmission\\ Rate} & \tabincell{c}{Content-aware\\ Transmission\\ Rate} \\ 
    \midrule
    NetInf \cite{gomez2012inferring}           & \tabincell{c}{\checkmark} &                                 &                                      & \tabincell{c}{\checkmark}                   &                                   &                                             &                                                           \\
    \midrule
    MultiTree \cite{gomez2012submodular}       & \tabincell{c}{\checkmark} &                                 &                                      & \tabincell{c}{\checkmark}                   &                                   &                                             &                                                           \\
    \midrule
    CONNIE \cite{CONNIE_myers2010convexity}    & \tabincell{c}{\checkmark} &                                 &                                      & \tabincell{c}{\checkmark}                   & \tabincell{c}{\checkmark}          &                                             &                                                           \\
    \midrule
    NetRate \cite{rodriguez2011uncovering}     & \tabincell{c}{\checkmark} &                                 &                                      & \tabincell{c}{\checkmark}                   & \tabincell{c}{\checkmark}          &                                             &                                                           \\                           
    \midrule
    InfoPath \cite{gomez2013structure}         & \tabincell{c}{\checkmark}  &                                 &                                      & \tabincell{c}{\checkmark}                   & \tabincell{c}{\checkmark}          & \tabincell{c}{\checkmark}                    &                                                           \\ 
    \midrule
    MoNet \cite{wang2012feature}               & \tabincell{c}{\checkmark} & \tabincell{c}{\checkmark}        &                                      & \tabincell{c}{\checkmark}                   & \tabincell{c}{\checkmark}          &                                             &                                                            \\
    \midrule
    MMRate \cite{wang2014mmrate}               & \tabincell{c}{\checkmark} & \tabincell{c}{\checkmark}        & \tabincell{c}{\checkmark}             & \tabincell{c}{\checkmark}                   & \tabincell{c}{\checkmark}          &                                             & \tabincell{c}{\checkmark}                                   \\
    \midrule
    DSTs \cite{xu2018contrastive}              & \tabincell{c}{\checkmark} & \tabincell{c}{\checkmark}        & \tabincell{c}{\checkmark}             & \tabincell{c}{\checkmark}                   &                                   &                                             &                                                            \\
    \midrule
    CSDK \cite{bourigault2014learning}         & \tabincell{c}{\checkmark} & \tabincell{c}{\checkmark}        & \tabincell{c}{\checkmark}             & \tabincell{c}{\checkmark}                   &                                   &                                             & \tabincell{c}{\checkmark}                                   \\
    \midrule
    ECM \cite{bourigault2016representation}    & \tabincell{c}{\checkmark} & \tabincell{c}{\checkmark}        & \tabincell{c}{\checkmark}             & \tabincell{c}{\checkmark}                   &                                   &                                             & \tabincell{c}{\checkmark}                                   \\
    \midrule
    IDEP \cite{gao2017novel}                   & \tabincell{c}{\checkmark} & \tabincell{c}{\checkmark}        & \tabincell{c}{\checkmark}             & \tabincell{c}{\checkmark}                   &                                   &                                             & \tabincell{c}{\checkmark}                                   \\
    \midrule
    DiffusionGAN \cite{DBLP:conf/ispa/0006ZZ019}   & \tabincell{c}{\checkmark} & \tabincell{c}{\checkmark}        & \tabincell{c}{\checkmark}             & \tabincell{c}{\checkmark}                   &   \tabincell{c}{\checkmark}        &                                             & \tabincell{c}{\checkmark}                                   \\
    \midrule
    HUGE \cite{DBLP:conf/ispa/0006ZZ019}   &                           & \tabincell{c}{\checkmark}        & \tabincell{c}{\checkmark}             & \tabincell{c}{\checkmark}                   &   \tabincell{c}{\checkmark}        &                                             & \tabincell{c}{\checkmark}                                   \\
    \bottomrule
    
  \end{tabular}
\end{table*}

\subsubsection{Analysis and Discussion}
Diffusion embedding models excellently integrate semantics of users and information. 
They can automatically determine the relevance of users and information. 
Compared with those time-series models that can only infer the past information propagation trajectory, diffusion embedding models can predict future propagation relationships because they mine relationships between users and information semantics.
However, the network topology factors have not been introduced into diffusion embedding models. 
It should be noted that cascade records inputted into diffusion embedding models must include sender and receiver because the diffusion embedding models need to arrange the users' location in the latent semantic space based on these observable forwarding relationships. In contrast, likelihood maximization models only need receivers. 
The summary of relationship inference models is shown in Table \ref{TAB:SummaryOfNetworkInference}.
The more data the model needs, the stronger the model's inference capacity is.

\section{Scenario-Specified Diffusion Modeling and Downstream Applications} 
\label{sec:ScenarioAndApplication}
\begin{figure}[H]
  \centering
  \includegraphics[width=5.6in]{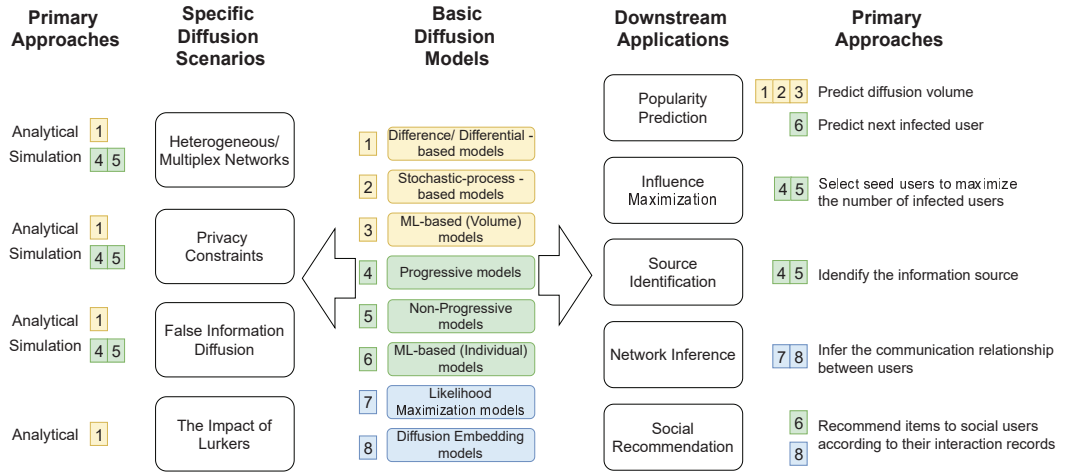}
  \caption{Specific diffusion scenarios and downstream applications based on these diffusion models.}
  \label{FIG:Applications}
\end{figure}

As shown in Fig. \ref{FIG:Applications}, in this section, we will discuss diffusion modeling in some specific scenarios, and downstream applications based on those universal diffusion models (\textit{i.e., they are not highly relevant to diffusion scenarios.}) introduced in Section \ref{sec:TaxonomyDiffusionModels}.

\subsection{Diffusion Modeling in Specified Scenarios}
\label{sec:SA:DiffusionScenario}

\subsubsection{Hetergeneous/Multiplex Networks}
In the real world, users have diverse social relationships, and they form multiplex/heterogeneous social networks.
For example, users can have Twitter and Facebook accounts, which means that the user is in two relatively isolated social circles at the same time. 
As shown in Fig. \ref{FIG:DiffusionScene}, there are intersections among different network layers, and information can be propagated across various networks via these intersections. 
Obviously, the information spread on different networks affects each other. 
When studying heterogeneous network information diffusion modeling, researchers usually assume that the network $G^{heter}$ is composed of $L$-layer networks $G^l$ with same nodes but different edges, denoted by $G^{heter}=\{G^1,G2,\ldots,G^L\}$, $G^l = (V, E^l)$. 
$E^l$ denotes the edges in the $l$-th layer of the network. 
The diffusion process consists of intra-layer diffusion and inter-layer diffusion. 

Universal models can be adapted in the following three manners to capture the diffusion dynamics in heterogeneous networks. 
One approach is the \textbf{extended cascade}. 
If the user $v$ is infected at $i$-th layer, he will try to infect his corresponding position in other networks with probability $\gamma$. 
In this way, a user remains susceptible unless he is not infected in all layers. 
Li \textit{et al.} \cite{PhysRevE.92.042810} described the intra-layer and inter-layer diffusion processes by bond percolation and cascading failure. 
And they proved that the inter-layer interactions could speed up the information diffusion process. 
Wang \textit{et al.} \cite{PhysRevE.96.032304} introduced the interplayer recovery mechanism, which allows users to be infected and recovered across layers. 
Experiments on synthetic networks show that the interplayer recovery mechanism can promote the diffusion process, and the optimal recovery rate can be calculated by deriving the diffusion equations. 
Zheng \textit{et al.} \cite{DBLP:journals/jpdc/ZhengXGD18} defined states for each layer of the user. 
The user's states at each moment can be expressed as a combination of all layer states, and a global transmission tree can be constructed according to combined user states. 
However, if there are too many layers, it will cause a state explosion. 
Sahneh \textit{et al.} \cite{sahneh2013generalized} employed first-order mean-field approximation to reduce state space from exponential to polynomial. 
Li \textit{et al.} \cite{DBLP:journals/tcss/LiWWX19} studied the diffusion of semantic-related information in multiplex networks. 
They added an ``acceleration'' factor $\alpha$ to indicate the competition or cooperation relationship among different information. 
It is shown that if two pieces of information are competitive, they will hinder each other during their diffusion process, and vice versa. 
Another approach is \textbf{extended threshold}. 
Zhang \textit{et at.} \cite{zhang2016information} assigned the activation threshold $\theta_u$ for each user $u$. 
The user's activation impact on each network will accumulate, and will be activated if the threshold is exceeded. 
A third approach is \textbf{coupling}. 
Nicosia \textit{et al.} \cite{PhysRevLett.118.138302} established connections among different layer diffusion processes through parameters correspondence. 
Specifically, the changes of user states in the $i$-th layer will lead to adjustments of activation function parameters in other layers. 
This mechanism can induce some collective phenomena such as spontaneous explosive synchronization and heterogeneous distribution of allocations. 

\subsubsection{Privacy Constraints}
In addition to structural characteristics, privacy settings also affect user diffusion capability. 
Generally, without forwarding, messages posted by Whatsup/Wechat users can only be browsed by their friends, while messages posted by Twitter users are accessible for almost everyone. 
Moreover, they can also control some other users' access by configuring privacy settings \cite{DBLP:conf/acsac/HuAJ11}. 

The main idea of existing privacy-aware diffusion models is to classify different users according to privacy settings, and define corresponding state transition rules. 
Zhu \textit{et al.} \cite{DBLP:journals/cj/ZhuHL15} separately defined the privacy policy of users and connections. 
They divide user privacy levels into three categories: public, private, and rigorous, representing access rights for everyone, only friends, and only themselves. 
And there are three types of relationships: bi-directional, semi-directional, and non-neighbor. 
Based on these assumptions, they proposed DMPS, an extended SI model. The key to DMPS is that users can choose to propagate information publicly or privately. 
Simulation results show that the privacy settings slow down the diffusion speed. 
Livio \textit{et al.} \cite{10.1007/978-3-319-54241-6_8} divides users into multiple different privacy classes, and the probability that a user $v$ belongs to level $j$ is denoted as $\beta_j \in \left[0,1\right]$. 
$\beta_j$ can be understood as the possibility for the user to participate in the diffusion process. 
In other words, it scales back the propagation capability in a privacy protection environment. 
Besides, George \textit{et al.} \cite{giakkoupis:hal-01184246} proposed a distributed algorithm, RIPOSTE, which achieves the social property of plausible deniability. 
It can infer information popularity and user attitudes through parameterized methods.

\subsubsection{False Information Diffusion}
False information refers to texts, images, and other contents that are meant to manipulate people in society, such as rumors, false news, hoax, and so on \cite{MEEL2020112986}. 
Their diffusion dynamics is a well-studied researched area, including \textit{diffusion dynamic analysis}, \textit{diffusion modeling}, \textit{containment}, \textit{source identification}, and \textit{detection}. 
In this scenario, infected users can become recovery during the interaction with other users \cite{daley_epidemics_1964}. 
For example, when two false information spreaders meet, they may lose interest in spreading the rumor any further, or stop spreading if fact-checkers debunk them. 

Many researchers have comprehensively analyzed the characteristics of false information diffusion, such as user behavior, network structure, and steady-state analysis. 
Vosoughi \textit{et al.} \cite{vosoughi2018the} found that fake news is more attractive due to its novelty, which leads to more forwarding. 
Even worse, false information can guide public perceptions \cite{10.1111/jcom.12284, 10.1111/j.1460-2466.2010.01497.x}. 
Wang \textit{et al.} \cite{Wang_2013} found that the trust mechanism reduces the maximum rumor influence and its diffusion speed. 
Jie \textit{et al.} \cite{JIE2016129} depicted the relationship of promotion, suppression, and independence among multiple rumors by defining interaction factors. 
Yang \textit{et al.} \cite{ZHOU2007458} discovered that the density of final infected users is related to the network topology, and random networks have higher infection densities than scale-free networks.  
In addition, researchers studied false information diffusion in many other specific situations, such as time-varying networks \cite{jiang2016rumor}, delay \cite{ZHU2016119}, behavior uncertainty \cite{ZHU201829}, noise interference \cite{ZHU2017750}, message evaluation \cite{SHIN2018278}. 

Generally, the mainstream approaches of diffusion modeling are first to define the \textbf{states and transition rules}, and then adopt \textbf{analytical} or \textbf{simulation} methods to capture false information diffusion dynamics. 
Similar to Epidemic models, \textbf{states and transition rules} are intend to describe the users behavior patterns in rumor diffusion, for example, SIS (Susceptible - Infected - Susceptible) \cite{PhysRevLett.86.3200}, SIR (Susceptible - Infected - Recovered) \cite{PhysRevE.64.066112}, SEIR (Susceptible-Exposed-Infected-Removed) \cite{XWYZ_huo2017dynamical}, SHIR (Susceptible - Infected - Hibernator - Removed) \cite{ZHAO20122444}. 
\textbf{Analytical} approaches are primarily mean-field difference/differential-based approaches, deriving equations to describe the proportion changes of different classes. 
\textbf{Simulation} approaches design progressive or non-progressive models according to the transition rules, conducting simulations of diffusion process based on the network topology. 

Containment of false information can be addressed by network obstruction or counterbalance. 
Network obstruction block critical social accounts, and counterbalances locate fact-checkers (\textit{i.e.,} they verify the correctness of information) or deliver correct information at critical locations \cite{NGUYEN20132133, 8194896}. 
Obviously, counterbalances are more moderate and effective \cite{YANG2020113}.
Tambuscio \textit{et al.} \cite{10.1145/2740908.2742572} discovered that the effectiveness of fact-checkers depends on the gullibility and forgetting probability. 
Further, in case of a low forgetting rate, network segregation can make misinformation speed faster. Otherwise, the effect is not apparent. 
However, in some cases, such as urban legends, the fact-checker mechanism does not seem to be effective \cite{tambuscio_fact-checking_2019}. 
Young \textit{et al.} \cite{doi:10.1177/1077699017710453} verified video is more effective in rectifying public misperceptions. 
Nekmat \textit{et al.} \cite{doi:10.1177/2056305119897322} thought that skeleton of mainstream media and fact-checking will promote the diffusion of non-mainstream media.

Source identification refers to inferring the source of rumors based on the results of rumors diffusion.
It is a downstream application based on diffusion models, which will be discussed in Section \ref{sec:SSDMADA:DA:SI}. 
Current detection mainly adopts machine learning or NLP approaches, which are beyond the scope of this article. 
Readers can refer to this literature \cite{10.1145/3161603, MEEL2020112986}.

\subsubsection{Impact of Lurkers on the Information Diffusion}
As shown in Fig. \ref{FIG:DiffusionScene}, the user's interactive behaviors promote information diffusion. 
However, 90\% of social network users are lurkers, who only browse or reshare contents posted by others and do not contribute to the community \cite{chiang_impact_2015}. 
Researchers have conducted research in various fields, including \textbf{lurker behaviors analysis}, \textbf{ranking of lurkers}, and \textbf{delurking}. 

For \textbf{lurker behaviors analysis}, most researchers aim to study the impact of lurker behaviors on information diffusion. 
Tagarelli \textit{et al.} \cite{tagarelli_time-aware_2015, tagarelli_lurking_2018} systematically analyzed the interactive behaviors of lurkers. 
They found that the temporal trends of lurkers' time-consuming actions present sharper shifts and more noisy clusters. 
And the average response period of lurkers seems to be at least twice that of active ones. 
Taking lurkers' behaviors into consideration, Fu \textit{et al.} \cite{fu_dynamic_2021} proposed SEAIR (Susceptible - Lurker - Super - Normal - Recovered) model to capture diffusion dynamics of super spreaders and lurkers. 

\textbf{Ranking of lurkers} methods are used to mining silent members in social networks. 
The topology-based analysis is an important method, which identifies lurkers based on the structural characteristics of users, such as the ratio of in/out-degrees \cite{tagarelli_lurking_2014}. 
Subsequently, they proposed varieties of ranking approaches, including time-aware ranking \cite{tagarelli_time-aware_2015}, learning-to-rank \cite{perna_learning_2018}, and multiplex network ranking \cite{perna_identifying_2018}. 

\textbf{Delurking} methods aim to engage lurkers in contributing via information propagation, which can be considered as the particular case of targeted influence maximization \cite{interdonato_community-based_2016, 10.1145/2808797.2809394}. 
Further, many researchers have explored some factors that affect the interactive behavior of lurkers, such as the structure of product-review network \cite{chiang_impact_2015}, and argument quality \cite{alarifi_posters_2015}. 
These factors are critical to discovering potential users in marketing.

\subsection{Downstream Applications}
\label{sec:DAOI:Applications}

\subsubsection{Popularity Prediction}
Popularity prediction is of practical significance in many areas. 
For example, the government can analyze the popularity of some sensitive information (\textit{e.g.}, rumors, terrorism information), and predict whether there will be an outbreak or extinction. 
As shown in Figure. \ref{FIG:Applications}, difference/differential-based, stochastic-process-based, and ML-based (Diffusion Volume and Individual Adoption) models are commonly used in popularity prediction. 
As these models are either diffusion volume models or data-driven individual adoption models, sophisticated individual interaction behaviors can be ignored. 
In addition, these models are easier to exhibit some principles of the diffusion process, such as K-SC \cite{yang2011patterns}, steady rates of change and interrupted by sudden bursts \cite{myers2014bursty}. 

If the other models are to be applied to predict popularity, there are many complicated problems. 
First, there is plenty of information diffusing in social networks, and the interrelation among these pieces of information should not be ignored. 
Second, accurate prediction requires a precise understanding of the user's re-sharing mechanisms (including user's cognition, herd behavior), which is a challenge of all time. 
Third, with the enhancement of privacy protection, it is difficult to obtain a complete network topology, and we also need to consider the network evolution during the information diffusion process.
Finally, we do not know how the external influence (from other platforms) affects the information diffusion on the observed platform. 
It means that the diffusion is almost restricted to the closed-world, far from the real world.

\subsubsection{Influence Maximization}
Given a social network topology \(G(V,E)\) and the diffusion model \(\mathcal{M}\), \(K\) users are selected from the user set \(V\) as the seed set \(\mathcal{I}(t)\), which maximize the expected number of active users at the end of the diffusion process.
IM has been a research hotspot since it was proposed because of its wide range of applications, such as marketing and public opinion control. 

Currently, solutions to the IM problem are primarily based on the greedy algorithm, as shown in Algorithm \ref{ALG:GREEDY}. 
The total round number is \(K\), and in \emph{i}-th round, the most influential inactive user \(u^{*}\) is moved to the seed set. 
Obviously, the accuracy and efficiency of the influence function, \(\sigma(\cdot)\), is the deterministic factor of the algorithm's efficiency. 
Existing IM algorithms mainly rely on time-series individual adoption models, more precisely, progressive models. 
Compared to the non-progressive models, their monotonicity and submodularity make them more efficient. 
However, \(\sigma(\cdot)\) costs a lot of resources. 
Kempe \cite{kempe2005influential} proved that \(\sigma(\cdot)\) based on IC/LT is \textit{\#P-hard}, and IM algorithms based on IC/LT are \textit{NP-hard}. 
The computing complexity will be higher if other individual-level models are adopted. 
Readers can refer to relevant literature \cite{li2018influencesurvey} for further understanding.  
\begin{algorithm}
  \caption{Greedy(\(K,\sigma\))\cite{li2018influencesurvey}}
  \label{ALG:GREEDY}
  \KwIn{\(K\): The number of seed users; \(\sigma(\cdot)\):Influence Function; $V$: User set}
  \KwOut{\(\mathcal{I}(0)\): Seed Set}
  \(\mathcal{I}(0)\leftarrow\oslash\)\;
  $i = 0$\;
  \While{$i<K$}{
    $u^{*}\leftarrow{\text{argmax}_{u\in{V\backslash{\mathcal{I}(0)}}}(\sigma(\mathcal{I}(0)\cup\left\{u\right\})-\sigma(\mathcal{I}(0)))}$\;
    \(\mathcal{I}(0)\leftarrow{\mathcal{I}(0)\cup{\left\{u^{*}\right\}}}\)\;
    $i=i+1$\;
  }
  
\end{algorithm}

\subsubsection{Source Identification}
\label{sec:SSDMADA:DA:SI}
Source identification tasks attempt to locate the origin of the information. It is different from the IP traceability in computer networks, which usually only identifies one of the many propagation participants. 
Moreover, it does not mean that the found one is the actual origin of the information. 
Source identification has very vital practical significance, especially for rumor source identification. 

Similar to IM, source identification techniques also rely on time-series individual adoption models.  
The solution consists of two steps. 
First, it measures user influence using time-series individual adoption models. 
Next, it finds the most appropriate users according to the metrics, such as Jordan centrality \cite{hage1995eccentricity}, and Eigenvector centrality \cite{bonacich1987power}. 
Deterministic factors of source identification are similar to influence maximization ones. 
Many researchers have done detailed investigations in this field \cite{jiang2017identifying, MEEL2020112986}, and we will not discuss them in details due to space constraints. 

\subsubsection{Network Inference}
Network inference tasks aim at inferring the propagation relationships in social networks. 
In addition, they can also be used to complete the network structure. 
Researchers often assume that information flows through edges between users in a social network. 
However, it is almost impossible to obtain a complete network topology due to privacy protection and network evolution. 
Relationship inference models can be used in network inference, and we summarize them in Table \ref{TAB:SummaryOfNetworkInference}. 

Since other models are ``many-to-many'' activation relationships, they can hardly be used directly to infer diffusion networks. 
Diffusion volume models do not involve individual interactions, so that they are not suitable for inferring diffusion networks. 
Individual adoption models can only determine the propagation relationship among \(\mathcal{I}(t)\) and \(\mathcal{I}(t+1)\). 
For example, in terms of the IC model, we can only know that an active user is activated by at least one of his neighbors, but cannot figure out a particular neighbor user.
Regarding the LT model, the propagation relationship is related to the input order. 
For instance, given user \(v\) with an activation threshold \(\theta_{v}=0.5\), if the threshold accumulated until time \(t\) is 0.4, the activation contributions of the two newly activated neighbors \(u_1\) and \(u_2\) are 0.1 respectively. 
Therefore, it is impossible to determine whether \(u_1\) or \(u_2\) is the actual activator.

\subsubsection{Social Recommendation} 
Social recommendation, or personalized recommendation, intends to recommend items that users may interest in based on their interaction records. 
Consequently, data-driven models are applicable for the social recommendations. 
Generally, \textit{filtering-based} and \textit{deep-learning-based} are primary approaches used in social recommendation. 
\textit{Filtering-based} approaches encode the social context (\textit{e.g.,} user-item interaction records, item attributes, and user attributes) into matrices. 
And then, they recommend items through filtering algorithms, such as matrix factorization or similarity calculation. 
\textit{Deep-learning-based} approaches generate the recommendation results through the end-to-end classifier \cite{KIM2010212}. 
There is enormous commercial potential in this area, and these models are comprehensively reviewed in literature \cite{KIM2010212, 272f266581f14ea7a0987d675bedb476,dhelim2021survey}.

\begin{figure}
  \centering
  \includegraphics[width=5.2in]{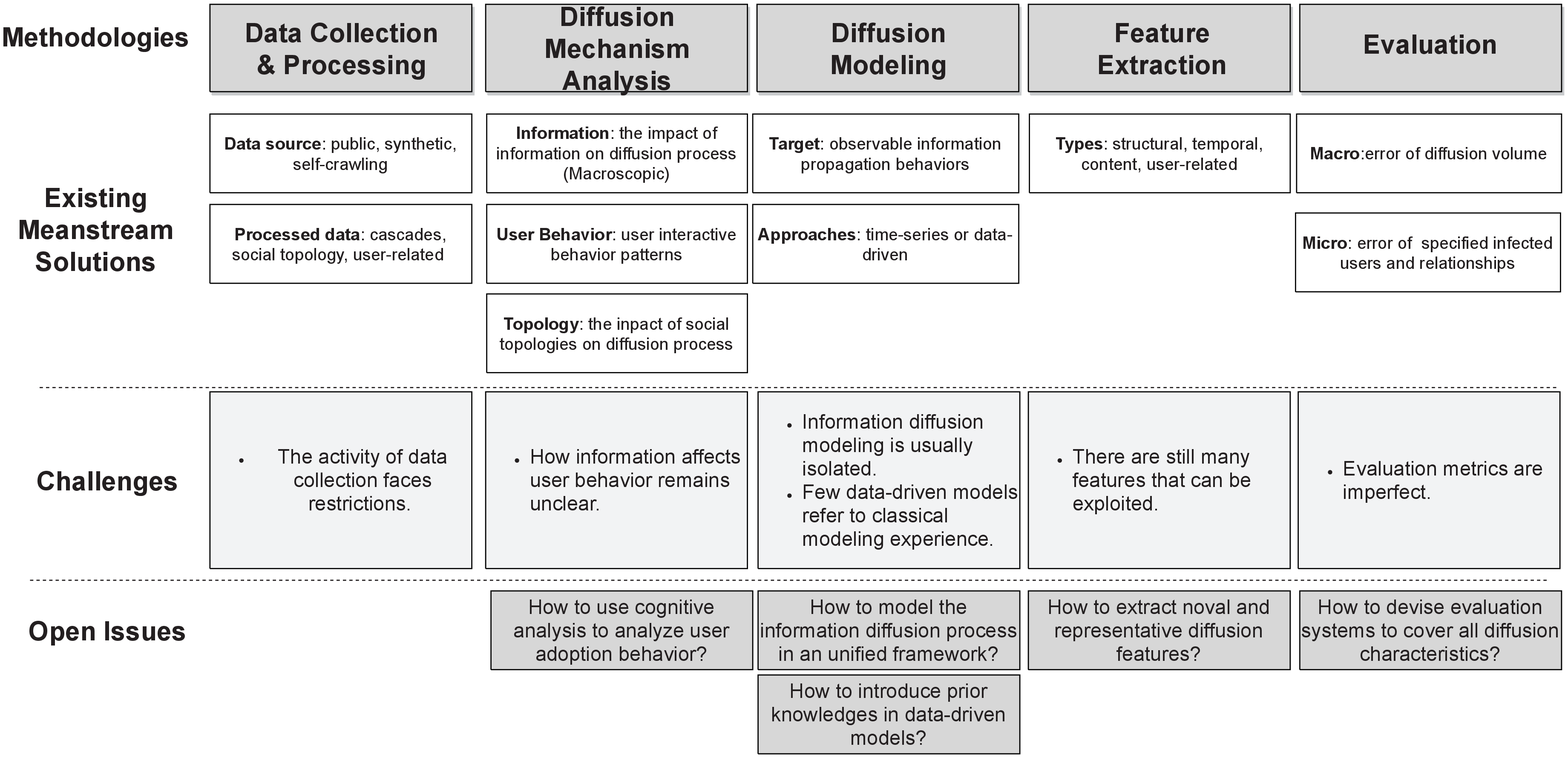}
  \caption{Open issues and challenges.}
  \label{FIG:Challenges}
\end{figure}

\section{Challenges and Open Issues}
\label{sec:Challenges}

In this section, challenges and open issues of diffusion modeling techniques are discussed corresponding to the methodology in Section \ref{sec:RMP:Methodology}. 
Fig. \ref{FIG:Challenges} is an overview of chanlleges and open issues. 

\emph{1) Data Collection \(\&\) Processing}

\textbf{Challenge :}\emph{The activity of data collection faces restrictions.}
Recently, crimes of user privacy (\textit{e.g.}, fraud, harassment, and personal information selling) make people and enterprises attach great concern over privacy protection. 
The enhancement of privacy protection is a gospel for ordinary users. However, it offers challenges for researchers. 
For example, the announcement of Twitter \cite{twitter_policy} says that it is not allowed to redistribute or syndicate Twitter datasets if they contain tweets. 
Facebook prevents access to the ID of users participating in public forums \cite{facebook_policy}. 
Although not many companies have performed relevant policies, it may block some researchers who lack data with the enhancement of privacy protection policy. 

\textbf{Solution:} Researchers can choose datasets from similar platforms to substitute, or crawl by themselves following platform policies.

\emph{2) Diffusion Mechanism Analysis}

\textbf{Challenge:}
\emph{How information affects user behavior remains unclear. }
Different users will show various behaviors when facing information. 
In early times, many classical models (\textit{e.g.}, SI \cite{SI_kermack1927contribution}, IC \cite{IC_goldenberg2001talk}, and LT\cite{LT_granovetter1978threshold}) think that all users are homogeneous, which means that the effect of information on each user is indistinguishable. 
These assumptions are far from the real world.
Subsequently, researchers assigned social attributes to each user make users exhibit various responses to the same information. 
Social attributes can be topological (\textit{e.g.}, friendship hops \cite{PDE_wang2012diffusive},social connectivity \cite{he2016cost}, and social role \cite{yang2015rain}) and user personal characteristics (\textit{e.g.}, user interests \cite{zhang2016retweet, hoang2017gpop}, user authority \cite{ma2012will, ma2013on}, and user identity \cite{zhang2013social,zhang2015influenced,liu2018c}). 
These personalized characteristics can be encoded into vectors or numerical values for propagation prediction. 
However, for a particular piece of information, how does it affect a user's social behavior? How strong will it affect everyone? How will a user modify the information? 
User's social actions depend on the match of their thinking and information semantics. 
These problems may require further study in combination with human cognition.

\textbf{Open Issue 1:}
\emph{How to analyze user adoption behavior in combination with human cognition?}

\emph{3) Diffusion Modeling}

\textbf{Challenge:}
\emph{Information diffusion modeling is usually isolated.} 
As shown in Fig. \ref{FIG:DiffusionScene}, user interactive behaviors are roughly divided into five parts in the information diffusion process, including observable and invisible actions. 
Current diffusion models almost only focus on the fourth part, information propagation, ignoring other behaviors. 
However, the information diffusion process is a complex system, and multiple behaviors affect information diffusion together. 
For example, the news is that a star was arrested for a crime, which may cause his fans to stop paying attention to him. 
Second, the user re-share behavior may cause him to be more intimate (distant) with a friend, or attract strangers to follow him. 
Researchers also have proven that information diffusion will drive the network evolution \cite{Weng:2013:RID:2487575.2487607}.
Third, users may append their comments to the tweet when forwarding it, which will shift the semantics of the original information. 
At present, some researchers have published their work in this field, such as the Time-driven model \cite{kempe2005influential} Correlated Cascades \cite{zarezade2017correlated}, Co-Evolve \cite{farajtabar2015coevolve}, and DynaDiffuse \cite{xie2015dynadiffuse}. 
However, these works seem only to introduce topological evolution.

\textbf{Open Issue 2:}
\emph{How to model the information diffusion process in a unified framework?}

\textbf{Challenge:}
\emph{Few data-driven models refer to time-series modeling experience. }
Time-series models design a series of mathematical expressions for diffusion scenarios. 
These achievements include user interaction status and transition rules \cite{PhysRevE.64.066112, XWYZ_huo2017dynamical, fu_dynamic_2021}, user behavior laws (\textit{e.g.,} delay \cite{saito2011learning, guille2012predictive}, distribution of user behaviors \cite{yu2015micro}, opinion adoption \cite{nayak2019smart}), the intensity of social event occurrence \cite{shen2014modeling, SEISMIC_zhao2015seismic}, periodicity \cite{SPIKEM_matsubara2012rise, Bao:2016:MPP:2983323.2983868, kobayashi2016tideh}, etc. 
The effectiveness of these mathematical models has been verified in multiple scenarios. 
Some articles involve these mechanisms \cite{cao2017deephawkes, DBLP:conf/icde/Chen0ZTZZ19, DBLP:conf/sigir/ChenZ0TZZ19, wu2018adversarial}. 
Although neural networks have strong approximation capabilities, we believe that introducing the knowledge of time-series techniques can help model optimization, such as reducing the search space. 
We believe that this knowledge can help improve the module architecture and the aggregation of module outputs. 

\textbf{Open Issue 3:}
\emph{How to introduce prior knowledge in data-driven models?}

\emph{4) Feature Extraction}

\textbf{Challenge:}
\emph{There are still many features that can be exploited to be discovered. }
For the moment, hundreds of four types of features are used for diffusion prediction. 
However, there are still some scopes that are not involved. 
First, current network structural features are mostly static features (\textit{e.g.}, local features \cite{zhang2013social,zhang2015influenced}, global features \cite{cao2020popularity}, and retweet graph \cite{cheng2014can}), but do not include dynamic features such as network evolution. 
Second, user-related features are mainly focused on individual-level features. 
However, with the in-depth research, researchers have discovered that user behavior becomes gradually similar as the interaction process progresses \cite{crandall2008feedback}. 
These all indicate that there are still many features that are worth being extracted. 
Representative features enable diffusion models to have a better performance. 

\textbf{Open Issue 4:}
\emph{How to extract novel and representative diffusion features?}

\emph{5) Evaluation}

\textbf{Challenge:}
\emph{Evaluation metrics are imperfect. }
Generally, the evaluation of these models is from both macroscopic and microscopic perspectives. 
Researchers use macroscopic metrics (\textit{e.g.}, RMSE, MAE, and MAPE) to verify the correctness of the overall diffusion volume, and use microscopic metrics (\textit{e.g.}, Precision, Recall, and F1-score) to evaluate the individual prediction accuracy. 
These evaluation indicators are not comprehensive. 
For example, few people consider the length of the propagation path.
Qin \textit{et al.} \cite{qin2017efficient} discovered that the cascade depth predicted using IC models is dozens of times greater than actual situations. 
From this survey, at least, stability (\textit{i.e.,} multi rounds experiments) and robustness (\textit{i.e.,} across multiple datasets) should be considered. 

\textbf{Open Issue 5:}
\emph{How to devise evaluation systems to cover all diffusion characteristics?}

%---------------------------------------------------------------
% SECTION 5 CONCLUSION
%---------------------------------------------------------------

\section{Conclusion}
\label{sec:Conclusion}

In this article, we presented a thorough review of information diffusion modeling. 
Firstly, we state basic notions and the methodology of diffusion modeling. 
Secondly, we categorized representative diffusion models into three classes and analyzed them from three perspectives: assumptions, methods, and pros and cons. 
Thirdly,  some scenario-specified diffusion modeling and downstream applications are summarized. 
Finally, conforming to modeling methodology, challenges and open issues are discussed. 

We think that diffusion mechanisms analysis and diffusion modeling deserve more attention. 
Moreover, both time-series and data-driven models should be valued, and they should promote each other to obtain better predictive capabilities.
Besides, it is meaningful to develop more comprehensive evaluation metrics to measure the performance of diffusion models. 
With the explosion of data resources and computing power, researchers should combine other disciplines such as NLP and neural science to conduct in-depth researches on information diffusion modeling.
Hopefully, this survey can be used as a reference and guideline for future research. 

\section{Acknowledgments}

This work was supported by the National Natural Science Foundation of China (Grant No.61902013, No.U1636208, No.61862008.) and the Beihang Youth Top Talent Support Program (Grant No.YWF-20-BJ-J-1038, No.YWF-21-BJ-J-1039.)

\bibliographystyle{ACM-Reference-Format}
\bibliography{sample-base}

\section{APPENDIX I}
\label{sec:Appendix}
\begin{table}[H]
  \caption{Structural Features}
  \label{TAB:FeaturesOfStructural}
  \centering
  \begin{tabular}{p{1.2cm} p{3cm} p{8cm}}
      \toprule
      \textbf{Types}       &\textbf{Features}  &  \textbf{Definations} \\
      \midrule
      Network   &$node\_degree$              & Node degree \cite{cheng2014can,DBLP:conf/sdm/DingLLJ18,kong2014predicting, gao2014effective} \\
      Network   &$graph\_density$              & Graph density \cite{ma2012will, ma2013on,kong2014predicting, gao2014effective} \\
      Network   &$edge\_weight$              & Edge weight \cite{ma2012will, ma2013on} \\
      Network   &$num\_broad\_user$              & The number of border users \cite{ma2012will, ma2013on, gao2014effective} \\
      Network   &$user\_connect$              & User connectivity (\textit{e.g.,} ratio between the number of connected components and the number of nodes) \cite{ma2012will, ma2013on} \\
      Network   &$clustering$              & Network clustering \cite{hoang2017gpop, gao2014effective} \\
      Network   &$reciprocity$              & Benefit and investment between users \cite{yuan2016will, gao2014effective} \\
      Network   &$node\_auth$               & Node authority \cite{ma2012will}\\
      Network   &$triangles$                & Users form triangles \cite{Vu:2011:DEM:3104482.3104590, ma2012will}\\
      Cascade   & $orig\_user\_struc$        & Structural features of the original user (\textit{e.g.}, degree) \cite{cheng2014can, DBLP:conf/sdm/DingLLJ18} \\
      Cascade   & $early\_user\_struc$        & Structural features of early adopters (\textit{e.g.}, degree) \cite{cheng2014can, DBLP:conf/sdm/DingLLJ18} \\
      Cascade   & $act\_connect$        & Connections between active users (\textit{i.e.,} $\left\{(v_i, v_j)|(v_i, v_j)\in E, 0<i<j\right\}$) \cite{cheng2014can} \\
      Cascade   & $depth$        & The depth of the retweet tree \cite{cheng2014can} \\ 
      Cascade   & $retw\_path$        & Retweet path of cascades \cite{cao2017deephawkes} \\ 
      Cascade   & $num\_hops$        & Number of hops in diffusion process \cite{yang2010predicting} \\
      Cascade   & $cascade\_graph$        & Diffusion cascade graph \cite{wang2017topo, DBLP:conf/icde/Chen0ZTZZ19} \\
      \bottomrule
  \end{tabular}
\end{table}

\begin{table}[H]
  \caption{Temporal Features}
  \label{TAB:FeaturesOfTemporal}
  \centering
  \begin{tabular}{p{2cm} p{3cm} p{8cm}}
      \toprule
      \textbf{Types}       &\textbf{Features}  &  \textbf{Definations} \\
      \midrule
      Sequential   & $int\_time$        & Time ticks of social interactions \cite{jiang2014fema, hoang2017gpop, 10.1145/1871437.1871691} \\
      Sequential   & $frequency$            & Tweet frequency \cite{suh2010want} \\
      Sequential   & $time\_interval$        & Time elapsed between the $i$-th and $j$-th reshare \cite{cheng2014can,DBLP:conf/mm/DingWW19, gao2014effective, hong2011predicting} \\
      Sequential   & $exp\_strength$   & Number of users who saw the tweet per unit time \cite{cheng2014can} \\
      Sequential   & $reshare\_strength$   & Number of users who reshare the tweet per unit time \cite{cheng2014can, ma2012will, ma2013on} \\
      Statiscical  & $change\_ratio$              & Change ratio of forwarding number in consecutive time windows \cite{tsur2012s}  \\
      Statiscical   & $act\_time\_dist$   & Distribution of active time \cite{cheng2014can, hong2011predicting} \\
      Statiscical   & $time\_shape$   & Shape of time series \cite{kong2014predicting} \\
      Statiscical   & $time\_decay$   & Time decay effect \cite{cao2017deephawkes, 10.1145/1871437.1871691} \\
      Statiscical   & $time\_arrive$   & Time taken for first $k$ retweet to arrive \cite{gao2014effective} \\
      \bottomrule
  \end{tabular}
\end{table}

\begin{table}[H]
  \caption{Content Features}
  \label{TAB:FeaturesOfContent}
  \centering
  \begin{tabular}{p{2cm} p{3cm} p{8cm}}
      \toprule
      \textbf{Types}       &\textbf{Features}  &  \textbf{Definations} \\
      \midrule
      Statistical   & $has\_caption$             & Whether the posted media with a caption \cite{cheng2014can, DBLP:conf/sdm/DingLLJ18} \\
      Statistical   & $language$                & The language of the tweet  \cite{cheng2014can} \\
      Statistical   & $orin\_type$                & The type of the origin post (\textit{e.g.}, web page) \cite{cheng2014can} \\
      Statistical   & $text\_count$                & The length of the text or caption \cite{cheng2014can, DBLP:conf/mm/DingWW19, tsur2012s} \\
      Statistical   & $num_hashtag$                & The count of hashtags \cite{DBLP:conf/mm/DingWW19, suh2010want, jiang2015message} \\
      Statistical   & $num_url$                & The count of URLs \cite{suh2010want, ma2012will, ma2013on} \\
      Statistical   & $retweet\_ratio$     & The ratio of views and reposts of the tweets posted by early reshare users \cite{cheng2014can,DBLP:conf/sdm/DingLLJ18}   \\
      Statiscical  & $num\_mention$           & Number of \@ specified users in the tweet \cite{suh2010want, ma2012will, ma2013on, jiang2015message}  \\
      Semantic  & $media\_topic$              & The possibility that the posted video or image having a specific topic \cite{cheng2014can,DBLP:conf/sdm/DingLLJ18, DBLP:conf/mm/DingWW19}  \\
      Semantic  & $emotion\_sentiment$           & Emotion or sentiment of the tweet \cite{cheng2014can, tsur2012s, kong2014predicting}  \\
      Semantic  & $topic\_pop$           & Virality (popularity) of topic or tweet \cite{hoang2016microblogging, }  \\
      Semantic   & $topic\_dist$                & Topic distribution of tweet content \cite{hong2011predicting} \\
      Semantic  & $text\_deep\_features$           & Deep text features extracted by deep learning models \cite{DBLP:conf/mm/DingWW19}  \\
      Semantic  & $media\_mul\_mod$           & Multi media features extracted by deep learning models \cite{DBLP:conf/www/XieZZPYHLC20, DBLP:conf/mm/DingWW19}  \\
      Semantic  & $content\_sim$           & Content similarity \cite{jiang2014fema, zhang2016retweet}  \\
      Semantic  & $self\_disclos$           & Level of self-disclosure of tweet content  \cite{yuan2016will}  \\
      Semantic   & $hashtag\_content$                & Hashtag content \cite{tsur2012s, ma2012will, ma2013on, kong2014predicting} \\
      Semantic   & $location$                & Location hashtag \cite{tsur2012s} \\

      \bottomrule
  \end{tabular}
\end{table} 

\begin{table}[H]
  \caption{User-related Features}
  \label{TAB:FeaturesOfUserRelated}
  \centering
  \begin{tabular}{p{2cm} p{3cm} p{8cm}}
      \toprule
      \textbf{Types}       &\textbf{Features}  &  \textbf{Definations} \\
      \midrule
      Static   & $orig\_user\_attr$        & Attributes of the original user (\textit{e.g.}, age, gender) \cite{cheng2014can} \\
      Static   & $early\_user\_attr$        & Attributes of early adopters (\textit{e.g.}, age, gender) \cite{cheng2014can, DBLP:conf/sdm/DingLLJ18} \\
      Static   & $orig\_user\_follower$   & Users who follows the original user \cite{cheng2014can, tsur2012s, kong2014predicting} \\
      Static   & $orig\_user\_followee$   & Users that the original user is following \cite{cheng2014can, tsur2012s} \\
      Static   &$user\_inf$              & User influence (\textit{e.g.,} the number of fans, interest) \cite{cao2017deephawkes} \\
      Static   &$user\_sim$              & User similarity \cite{jiang2014fema, zhang2016retweet, 10.1145/1871437.1871691} \\
      Static   &$user\_prefer$              & User preference \cite{10.1145/1871437.1871691} \\
      Static   &$user\_auth$              & User authority \cite{ma2012will} \\
      Dynamic   &$user\_suscep$              & Susceptibility of users to be infected  \cite{hoang2016microblogging} \\
      Dynamic   & $orig\_user\_status$   & User social activity (\textit{i.e.} tweets published by the user) \cite{suh2010want,kong2014predicting} \\
      Dynamic   & $orig\_retweet\_ratio$              & The ratio of views and reposts of the tweets posted by the original user \cite{cheng2014can,tsur2012s, hong2011predicting,10.1145/2396761.2398634}  \\
      Dynamic   & $early\_retweet\_ratio$              & The ratio of views and reposts of the tweets posted by early adopters \cite{cheng2014can,DBLP:conf/sdm/DingLLJ18, 10.1145/2396761.2398634}   \\
      Dynamic   &$responsiveness$              & Responsiveness of the user to tweets  \cite{yuan2016will} \\
      \bottomrule
  \end{tabular}
\end{table} 

\end{document}